\documentclass{SciPost}

\hypersetup{
    colorlinks,
    linkcolor={red!50!black},
    citecolor={blue!50!black},
    urlcolor={blue!80!black}
}
\usepackage[bitstream-charter]{mathdesign}
\DeclareSymbolFont{usualmathcal}{OMS}{cmsy}{m}{n}
\DeclareSymbolFontAlphabet{\mathcal}{usualmathcal}

\fancypagestyle{SPstyle}{
\fancyhf{}
\lhead{\colorbox{scipostblue}{\bf \color{white} ~SciPost Physics }}
\rhead{{\bf \color{scipostdeepblue} ~Submission }}

\fancyfoot[C]{\textbf{\thepage}}
}

\usepackage{orcidlink}
\usepackage[nameinlink]{cleveref}
\usepackage{acronym}

\usepackage{feynmp}
\newacro{IR}[IR]{infrared}
\newacro{RG}[RG]{renormalisation group}

\newcommand{\eg}{e.g.}
\newcommand{\ie}{i.e.}
\newcommand{\fg}{\ensuremath{f_g}}
\newcommand{\fdsq}{\ensuremath{f_{\partial\partial\phi}}}
\newcommand{\fdd}{\ensuremath{f_{\partial\phi\partial\phi}}}
\newcommand{\fgk}{\ensuremath{f_{g,k}}}

\newcommand{\fgFP}{\ensuremath{f_{g\ast}}}
\newcommand{\fdsqFP}{\ensuremath{f_{\partial\partial\phi\ast}}}
\newcommand{\fddFP}{\ensuremath{f_{\partial\phi\partial\phi\ast}}}
\newcommand{\Fg}{\ensuremath{F_g}}
\newcommand{\Fdsq}{\ensuremath{F_{\partial\partial\phi}}}
\newcommand{\Fdd}{\ensuremath{F_{\partial\phi\partial\phi}}}
\newcommand{\Fgk}{\ensuremath{F_{g,k}}}
\newcommand{\Fdsqk}{\ensuremath{F_{\partial\partial\phi,k}}}
\newcommand{\Fddk}{\ensuremath{F_{\partial\phi\partial\phi,k}}}

\newcommand{\phsq}{\ensuremath{\left[p_h^2\right]}}
\newcommand{\ponesq}{\ensuremath{\left[p_{\phi_1}^2\right]}}
\newcommand{\ptwosq}{\ensuremath{\left[p_{\phi_2}^2\right]}}
\newcommand{\phsqdimless}{\ensuremath{\left[\tilde{p}_h^2\right]}}
\newcommand{\ponesqdimless}{\ensuremath{\left[\tilde{p}_{\phi_1}^2\right]}}
\newcommand{\ptwosqdimless}{\ensuremath{\left[\tilde{p}_{\phi_2}^2\right]}}

\newcommand{\argsofF}{\ensuremath{\left(\phsq,\ponesq,\ptwosq\right)}}
\newcommand{\argsoff}{\ensuremath{\left(\phsqdimless,\ponesqdimless,\ptwosqdimless\right)}}

\DeclareRobustCommand{\reflP}{\text{\reflectbox{P}}}

\makeatletter
\AddToHook{cmd/appendix/before}{\def\cref@section@alias{appendix}\def\cref@subsection@alias{appendix}}
\makeatother

\allowdisplaybreaks

\begin{document}

\pagestyle{SPstyle}

\begin{center}{\Large \textbf{\color{scipostdeepblue}{
%%%%%%%%%% TODO: Write your article's title here
Scalar-scalar-graviton vertex from Asymptotic Safety
}}}\end{center}

\begin{center}\textbf{
Tim Eckert\textsuperscript{1$\star$}\orcidlink{0009-0009-1791-0063} and
Benjamin Knorr\textsuperscript{2$\dagger$}\orcidlink{0000-0001-6700-6501}
}\end{center}

\begin{center}
{\bf 1} Theoretisch-Physikalisches Institut, Abbe Center of Photonics,\\ Friedrich-Schiller-Universit\"at Jena, Max-Wien-Platz 1, 07743 Jena, Germany
\\
{\bf 2} Institut f\"ur Theoretische Physik, Universit\"at Heidelberg, \\ Philosophenweg 12, 69120 Heidelberg, Germany
\\[\baselineskip]
$\star$ \href{mailto:tim.eckert@uni-jena.de}{\small tim.eckert@uni-jena.de}\,,\quad
$\dagger$ \href{mailto:knorr@thphys.uni-heidelberg.de}{\small knorr@thphys.uni-heidelberg.de}
\end{center}

\section*{\color{scipostdeepblue}{Abstract}}
\textbf{\boldmath{%
The unitarity and causality properties of asymptotically safe quantum gravity are still unsettled. Scattering amplitudes provide a direct test, but their computation requires fully momentum-dependent correlation functions. In this light, we compute the full scalar-scalar-graviton vertex from Asymptotic Safety within a mostly analytic approximation. Qualitatively, we find the expected logarithmic branch-cut structure, and we correctly recover the double-eikonal soft divergences predicted by Weinberg's soft theorem. Quantitatively, we find that non-minimal operators are generally as important as the minimal vertex stemming from the scalar kinetic term.
}}

\vspace{\baselineskip}

%%%%%%%%%% BLOCK: Copyright information
% This block will be filled during the proof stage, and finilized just before publication.
% It exists here only as a placeholder, and should not be modified by authors.
\noindent\textcolor{white!90!black}{%
\fbox{\parbox{0.975\linewidth}{%
\textcolor{white!40!black}{\begin{tabular}{lr}%
  \begin{minipage}{0.6\textwidth}%
    {\small Copyright attribution to authors. \newline
    This work is a submission to SciPost Physics. \newline
    License information to appear upon publication. \newline
    Publication information to appear upon publication.}
  \end{minipage} & \begin{minipage}{0.4\textwidth}
    {\small Received Date \newline Accepted Date \newline Published Date}%
  \end{minipage}
\end{tabular}}
}}
}
%%%%%%%%%% BLOCK: Copyright information

% For convenience during refereeing we turn on line numbers:
% \linenumbers

\clearpage

\vspace{10pt}
\noindent\rule{\textwidth}{1pt}
\tableofcontents
\noindent\rule{\textwidth}{1pt}
\vspace{10pt}

\clearpage

\section{Introduction}\label{sec:intro}

In recent years, Asymptotically Safe Quantum Gravity~\cite{Weinberg:1980gg, Reuter:1996cp} has matured to a serious contender for a theory of quantum gravity. It is based on an interacting \ac{RG} fixed point, whose scale symmetry provides a mechanism that solves the loss of predictivity which a perturbative quantisation of General Relativity faces. By now, the existence of a suitable fixed point -- the Reuter fixed point -- has been supported by substantial evidence, see the chapters~\cite{Knorr:2022dsx, Eichhorn:2022gku, Morris:2022btf, Wetterich:2022ncl, Martini:2022sll, Saueressig:2023irs, Pawlowski:2023gym, Platania:2023srt, Bonanno:2024xne} of the Handbook of Quantum Gravity~\cite{Bambi:2023jiz}, and~\cite{Bonanno:2020bil, Pawlowski:2020qer, Reichert:2020mja, Basile:2024oms, Eichhorn:2026uqj} for recent reviews.

The existence of a fixed point alone however does not guarantee other essential properties of a physical theory, like unitarity or causality. To properly assess whether Asymptotically Safe Quantum Gravity possesses these properties, a computation of scattering amplitudes and cross sections from first principles is highly informative. This necessitates the computation of fully momentum-dependent, Lorentzian correlation functions. The precise correlators that have to be computed for a given scattering event can be fully classified~\cite{Draper:2020knh, Knorr:2022lzn}, so that the main challenge is to compute them in practice from first principles.

Recent work in this direction has developed along two different lines: the first route computes such correlation functions in Euclidean signature, which are then Wick-rotated. The advantage is that these computations are based on decades of experience in resolving Euclidean \ac{RG} flows. However, they necessarily come with specific assumptions on the analyticity properties to extract real-time quantities. Along this route, first results were obtained numerically~\cite{Bonanno:2021squ, Pastor-Gutierrez:2024sbt, Chiesa:2026tlz}, followed by more recent analytical advances~\cite{Knorr:2026vax, Knorr:2026jcg}.

The second route performs computations directly in Lorentzian signature, by implementing a spectral version of the \ac{RG}~\cite{Horak:2020eng, Horak:2021pfr}. Technically, this is based on particular choices for the regularisation of the theory that preserve the causality even at finite cutoff. This approach has found first applications to Asymptotically Safe Quantum Gravity in~\cite{Fehre:2021eob, Kher:2025rve, Pawlowski:2025etp, Assant:2026dca}.

Both approaches have been successful in the computation of the graviton propagator and different interaction vertices, so that more and more ingredients of the full scattering amplitude are known or in reach.

In this paper, we contribute significantly to this effort of deriving asymptotically safe scattering amplitudes from first principles, by computing all channels of the momentum-dependent scalar-scalar-graviton vertex. A numerical approximation of one tensor structure of this vertex has been computed in~\cite{Chiesa:2026tlz}, where a specific momentum configuration was used that allowed to resolve two out of three dependencies. In this work, by leveraging the insights of~\cite{Knorr:2026vax, Knorr:2026jcg} to compute analytic approximations of correlation functions, we are able to give a largely analytic form of all channels of this vertex, including the full momentum dependence. In particular, for the first time in the context of asymptotically safe quantum gravity, we reproduce the double-eikonal soft divergences related to Weinberg's soft theorem exactly.

This paper is structured as follows. In \Cref{sec:scatt}, we motivate the study of the scalar-scalar-graviton vertex and scattering amplitudes in Asymptotically Safe Quantum Gravity. In \Cref{sec:asqg}, we describe our setup and the technical tools used in the computation. In \Cref{sec:results}, we present our results for the fixed-point and infrared form factors, the connection to Weinberg's soft theorem, and the limitations of our approximation. Finally, in \Cref{sec:conclusions}, we summarise our main results and discuss possible directions for future work.

\section{Gravity-mediated scalar scattering and soft gravitons}
\label{sec:scatt}

In this section, we briefly discuss some general aspects of the scattering amplitude for two scalars scattering into two other scalars in Minkowski space, including gravitational contributions and soft effects.

Consider two massless scalar fields $\phi$ and $\chi$ coupled to gravity.\footnote{Considering two distinct instead of a single scalar field allows us to focus the discussion on a single channel, which simplifies the exposition.} In the absence of \ac{IR} effects which we discuss in more detail below, the effective action that parameterises the process $\phi\phi\mapsto\chi\chi$ in a Minkowski background can be written as
\begin{equation}\label{eq:EA_FF}
\begin{aligned}
    \Gamma \simeq \int \text{d}^4x \, \sqrt{-g} \, \Bigg[ &\frac{1}{16\pi G_N} \left( R + \frac{1}{6} R F_R(\Box)R - \frac{1}{2} C^{\mu\nu\rho\sigma} F_C(\Box) C_{\mu\nu\rho\sigma} \right) \\
    & + \frac{1}{2} \phi F_\phi(\Box)\phi+ F_{R\phi\phi}(\Box_1,\Box_2,\Box_3) R\phi\phi + F_{Ric\phi\phi}(\Box_1,\Box_2,\Box_3)R^{\mu\nu} D_\mu \phi D_\nu \phi \\
    &+ (\phi\mapsto\chi) + F_{\phi\phi\chi\chi}(\{-D_i\cdot D_j\}_{i<j})\phi \phi\chi\chi \Bigg] \, .
\end{aligned}
\end{equation}
In this, $G_N$ is Newton's constant, the form factors $F_{R,C}$ (together with the Einstein-Hilbert term) parameterise the flat graviton propagator, $F_\phi$ parameterises the $\phi$-propagator, $F_{R\phi\phi,Ric\phi\phi}$ parameterise non-minimal interactions of the scalar $\phi$ with gravity, and $F_{\phi\phi\chi\chi}$ is the scalar self-interaction. Moreover, $\Box=-D^2$ with $D$ being the covariant derivative, and subscripts indicate the field that different operators act on -- an index $i$ indicates that the operator acts on the $i$-th field. We employ a mostly-minus convention for the metric $g$ and neglect the cosmological constant.

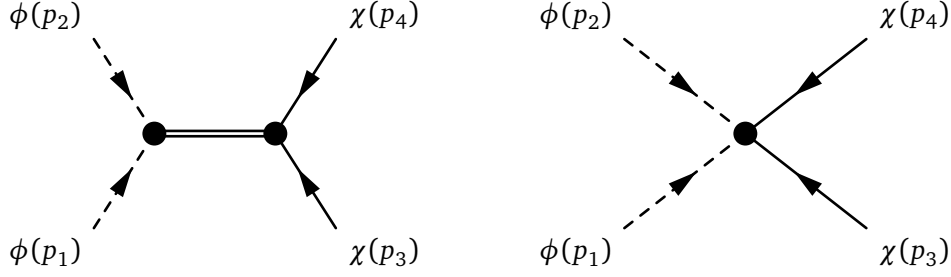
\begin{figure}
\vspace{0.2cm}
\centering
\unitlength = 1mm
\begin{fmffile}{mediated}
\begin{fmfgraph*}(40,25)
\fmfleft{i1,i2}
\fmfright{o1,o2}
\fmflabel{$\phi(p_1)$}{i1}
\fmflabel{$\phi(p_2)$}{i2}
\fmflabel{$\chi(p_3)$}{o1}
\fmflabel{$\chi(p_4)$}{o2}
\fmfv{decor.shape=circle,decor.filled=full,
decor.size=4thick}{v1,v2}
\fmf{scalar}{i1,v1}
\fmf{scalar}{i2,v1}
\fmf{fermion}{o1,v2}
\fmf{fermion}{o2,v2}
\fmf{double}{v1,v2}
\end{fmfgraph*}
\end{fmffile}
\hspace{2.5cm}
\begin{fmffile}{contactterm}
\begin{fmfgraph*}(40,25)
\fmfleft{i1,i2}
\fmfright{o1,o2}
\fmflabel{$\phi(p_1)$}{i1}
\fmflabel{$\phi(p_2)$}{i2}
\fmflabel{$\chi(p_3)$}{o1}
\fmflabel{$\chi(p_4)$}{o2}
\fmfv{decor.shape=circle,decor.filled=full,
decor.size=4thick}{v1}
\fmf{scalar}{i1,v1}
\fmf{scalar}{i2,v1}
\fmf{fermion}{o1,v1}
\fmf{fermion}{o2,v1}
\end{fmfgraph*}
\end{fmffile}

\vspace{0.5cm}
\caption{\label{fig:amp}The two diagrams that contribute to the exclusive scattering process $\phi(p_1)\phi(p_2)\to\chi(p_3)\chi(p_4)$: gravity-mediated diagram (left) and contact term (right). The double line in the left diagram indicates a fully dressed graviton propagator, whereas the dots represent fully dressed vertices. Since we are using the effective action to derive correlation functions, the tree-level diagrams contain the full information on the exclusive process.}
\end{figure}

The scattering amplitude for the exclusive process following from this action reads~\cite{Draper:2020knh}
\begin{equation}\label{eq:general_amp}
\begin{aligned}
    \mathcal A_\text{med} &= \frac{4\pi}{3} G_N \Bigg[ (s^2-6tu)\frac{(1-s F_{Ric\phi\phi}(s,0,0))(1-s F_{Ric\chi\chi}(s,0,0))}{s(1+s F_C(s))} \\
    &\quad - s^2 \frac{(1+2s F_{Ric\phi\phi}(s,0,0)-12 F_{R\phi\phi}(s,0,0))(1+2s F_{Ric\chi\chi}(s,0,0)-12 F_{R\chi\chi}(s,0,0))}{s(1+s F_R(s))} \Bigg] \, , \\
    \mathcal A_\text{cont} &= F_{\phi\phi\chi\chi}\left(\frac{s}{2},\frac{t}{2},\frac{u}{2},\frac{u}{2},\frac{t}{2},\frac{s}{2}\right) + \text{(sym)} \, ,
\end{aligned}
\end{equation}
where the subscripts med and cont indicate the contributions from the mediated (left panel of \Cref{fig:amp}) and the contact (right panel of \Cref{fig:amp}) diagram, respectively, and (sym) indicates the symmetrisation of the arguments as dictated by taking functional derivatives of the action. We also used the standard Mandelstam variables
\begin{equation}
    s = (p_{\phi_1}+p_{\phi_2})^2 \, , \quad t = (p_{\phi_1}+p_{\chi_1})^2 \, , \quad u = (p_{\phi_1} + p_{\chi_2})^2 \, .
\end{equation}
We have assumed the on-shell condition $p^2=0$ for the external scalars, so that both are massless.

The scattering amplitude above gives rise to well-known \ac{IR} effects~\cite{Weinberg:1965nx}. The underlying issue is that freely propagating particles are not asymptotic states in a theory with long-range interactions mediated by massless carriers like gravitons. Intuitively, the reason is that the exclusive two-to-two process cannot be distinguished from a similar process with additional, arbitrarily soft gravitons being emitted. Consequently, a computation of the above form factors will invariably lead to \ac{IR} singularities that, to leading order, are quadratic and linear in the logarithm of the external momentum, $\sim(\ln p_{\phi,\chi}^2)^2, \ln p_{\phi,\chi}^2$. The quadratic logarithms need to cancel in the amplitude when mediated and contact diagrams are summed.

There are different possibilities to handle this issue in order to arrive at finite predictions, and we briefly mention two of them. One is to consider inclusive cross sections, \ie{} one additionally considers the processes $\phi\phi\to\chi\chi h^n$ for all $n$. Combining these real-emission channels for gravitons with momenta below a detector resolution threshold with the virtual terms from the exclusive process, one obtains a finite result which however still depends on the physical resolution scale~\cite{Weinberg:1965nx, Kinoshita:1962ur, Lee:1964is}. This is the same principle applied to data from colliders like the LHC~\cite{Sterman:1977wj}, since not all of the outgoing particles are always detected in practice. A second possibility is to consider properly dressed (gauge-invariant) asymptotic states \`a la Faddeev-Kulish~\cite{Kulish:1970ut, Ware:2013zja}. Effectively, instead of a single scalar, the asymptotic state is a scalar surrounded by a ``graviton cloud'', incorporating the long-range gravitational field.\footnote{We note that the applicability of the Faddeev-Kulish construction to gravity is under debate, see \eg{}~\cite{Prabhu:2022zcr}.} This is a very natural viewpoint from the perspective of asymptotic symmetries, BMS charges, and the associated superselection rules~\cite{Choi:2017bna, Choi:2017ylo}, see also~\cite{Strominger:2017zoo} for topical lecture notes.

For our work, either of these solutions presents us with severe complications: either, we would need to resolve correlation functions with an arbitrary number of soft gravitons, or we would have to construct properly dressed asymptotic states. For this reason, as a first step here we only focus on the exclusive process described by \eqref{eq:general_amp} and a check of the ensuing soft factor. More concretely, in this work, we provide a first-principle computation of the full scalar-scalar-graviton vertex, including all tensor structures and the full momentum dependence. In the form factor language above, we compute the functions $F_{R\phi\phi,Ric\phi\phi}$. We thereby significantly improve the understanding of this scattering amplitude. This follows a recent initiative of classifying~\cite{Draper:2020bop, Draper:2020knh, Knorr:2022lzn} and computing~\cite{Christiansen:2014raa, Knorr:2021niv, Bonanno:2021squ, Fehre:2021eob, Pastor-Gutierrez:2024sbt, Kher:2025rve, Pawlowski:2025etp, Knorr:2026vax, Chiesa:2026tlz, Knorr:2026jcg, Assant:2026dca} the ingredients for scattering amplitudes within Asymptotic Safety, which we discuss in more detail in the next section.

\section{Asymptotically Safe Quantum Gravity}
\label{sec:asqg}

Asymptotically Safe Quantum Gravity is built on the hypothesis that quantum gravity can be formulated consistently as a quantum field theory. Underlying this is an interacting fixed point of the \ac{RG} flow, whose relevance properties provide a direct solution to the loss of predictivity encountered in the perturbative quantisation.

A key question in this approach to quantum gravity is whether the theory is unitary and causal, for a general discussion of these properties in quantum gravity, see~\cite{Platania:2022gtt}. A fixed point alone, while necessary to define the theory, in general does not guarantee these properties. This can be understood intuitively in terms of the amplitude: a fixed point would suggest bounded scattering amplitudes due to the scale invariance at high energies, but the partial wave amplitudes could still exceed unity, and thus violate unitarity.

While a non-perturbative proof of unitarity and causality seems out of reach for now, indications have been found in the affirmative. For example, the graviton propagator and spectral function have been computed in a flat background in~\cite{Christiansen:2014raa, Bonanno:2021squ, Knorr:2021niv, Fehre:2021eob, Pawlowski:2025etp, Knorr:2026jcg, Assant:2026dca}. This corresponds to a resolution of the form factors $F_{R,C}$ in \eqref{eq:EA_FF}. It was found that the graviton propagator only admits a single, massless pole. Moreover, the spectral functions of the on-shell spin two and the off-shell spin zero modes behave as expected from a classical graviton dressed by quantum fluctuations. These findings support the idea that the theory is unitary. Additionally, the propagator has been found to be analytic in the upper momentum half-plane, which via the Kramers-Kronig relations is evidence for a causal theory. In~\cite{Knorr:2026jcg}, the mechanism that removes spurious poles from the propagator which appear in derivative expansions has been identified -- the residue of these poles go to zero upon increasing the truncation order. This mechanism has been proposed first in~\cite{Platania:2020knd}.

More positive indications come from first computations of scattering amplitudes~\cite{Pastor-Gutierrez:2024sbt, Chiesa:2026tlz}. Results so far point towards well-behaved scattering cross sections. First promising results for spectral functions in gravity-matter systems have been put forward in~\cite{Kher:2025rve}.

A different route towards assessing unitarity is via probing positivity bounds~\cite{deRham:2022hpx}. While these traditionally only apply to effective field theories where massive degrees of freedom have been integrated out, their extension to gravity has been discussed more recently, see \eg{}~\cite{Bellazzini:2019xts, CarrilloGonzalez:2023cbf, Alberte:2020jsk, Herrero-Valea:2020wxz, Henriksson:2022oeu, Herrero-Valea:2022lfd, Bittar:2024xuc, Caron-Huot:2021rmr, Caron-Huot:2022ugt}. In the context of Asymptotic Safety, unitarity bounds for a photon-graviton system have been discussed in~\cite{Knorr:2024yiu, Eichhorn:2024wba}.

Our work takes a significant step towards the bigger goal of assessing unitarity and causality of Asymptotically Safe Quantum Gravity by providing one more ingredient of the scattering amplitude \eqref{eq:general_amp}. At the technical level, we compute the full scalar-scalar-graviton vertex in a Minkowski background.

\subsection{Functional renormalisation group}

The most common investigation strategy of Asymptotic Safety is based on the functional \ac{RG}. Instead of performing the path integral in a single step, momentum modes are integrated out shell by shell in a Wilsonian sense. For this, an \ac{IR} regulator is introduced which cuts off modes with momenta smaller than the cutoff scale $k$. With this at hand, an interpolating action $\Gamma_k$ can be introduced that connects to the microscopic action $S$ as $k\to\infty$, and to the standard effective action $\Gamma$ as $k\to0$.

In gauge theories as well as gravity, the non-linear structure of the symmetry transformation as well as the need to gauge-fix make it necessary to use the background-field formalism. In gravity, we split the metric $g$ into an arbitrary background metric $\bar g$ and fluctuations $h$ such that
\begin{equation}
    g_{\mu\nu} = \bar g_{\mu\nu} + h_{\mu\nu} \, .
\end{equation}
Consequently, the interpolating action $\Gamma_k$ is a two-field functional -- it not only depends on the diffeomorphism-invariant combination $\bar g+h$, but also additionally on $\bar g$ alone through the regularisation and gauge fixing. Physical diffeomorphism symmetry is then encoded in modified Nielsen and Slavnov-Taylor identities. The consequences for the flow have been discussed at length in the literature, for a review see~\cite{Pawlowski:2020qer}, and for a recent systematic solution of the issue using field redefinitions see~\cite{Ihssen:2026ucr}.

The functional $\Gamma_k$ fulfils the formally exact flow equation~\cite{Wetterich:1992yh, Morris:1993qb, Ellwanger:1993mw}
\begin{equation}\label{eq:wettericheq}
    k \partial_k \Gamma_k[\bar g,\bar g+h] = \frac{1}{2} \text{Tr} \left[ \left( \Gamma_k^{(0,2)}[\bar g,\bar g+h] + \mathfrak R_k[\bar g]  \right)^{-1} k \partial_k \mathfrak R_k[\bar g] \right] \, .
\end{equation}
In this, $\mathfrak R_k$ is the \ac{IR} regulator, and $\Gamma_k^{(0,2)}$ denotes the second functional derivative with respect to the full metric. Moreover, the trace indicates an integration over momenta and a sum over fields and indices. For a review of the functional \ac{RG}, see~\cite{Dupuis:2020fhh}. In the following, we will also often use the \ac{RG} ``time'' $t=\ln k/k_0$, where $k_0$ is an arbitrary reference scale, and an overdot will indicate a derivative with respect to $t$.

The flow equation \eqref{eq:wettericheq} is based on a Euclidean signature. While more recently, Lorentzian formulations have been proposed~\cite{Manrique:2011jc, Bonanno:2021squ, Fehre:2021eob, Banerjee:2022xvi, Saueressig:2023tfy, DAngelo:2023wje, DAngelo:2023tis, Ferrero:2024rvi, Banerjee:2024tap, Thiemann:2024vjx, Pastor-Gutierrez:2024sbt, Saueressig:2025ypi, Kher:2025rve, DAngelo:2025yoy, Pawlowski:2025etp, Assant:2026dca}, we will solve the Euclidean equation and perform a Wick rotation afterwards.

\subsection{Setup}

Our goal is to compute the scalar-scalar-graviton vertex by solving \eqref{eq:wettericheq} in an approximation that we are describing next. For simplicity, we only consider a single, shift-symmetric scalar field $\phi$. Consequently, we are interested in taking two scalar, and one graviton derivative of \eqref{eq:wettericheq}.

We start by describing the ``minimal'' ingredients, that is everything that is not related to the non-minimal form factors $F_{R\phi\phi,Ric\phi\phi}$. In this sector, we approximate
\begin{equation}\label{eq:gamma_minimal}
    \Gamma_k^\text{minimal} \approx \int \text{d}^4x\, \sqrt{g} \, \left[ -\frac{R}{16\pi G_{N,k}} + \frac{1}{2} D^\mu\phi D_\mu\phi \right] \, .
\end{equation}
Newton's constant $G_{N,k}$ now depends on the fiducial \ac{RG} scale $k$. For this work, we neglect the form factors $F_{R,C}$ and approximate the scalar form factor $F_\phi$ by the classical kinetic term. The gravitational sector is completed by a standard linear gauge fixing,
\begin{equation}
    \mathcal F_\alpha = \left( \delta_\alpha^{\phantom{\alpha}\mu} \bar D^\nu - \frac{1+\beta}{4} \bar g^{\mu\nu} \bar D_\alpha \right) h_{\mu\nu} \, , \qquad S_\text{gf} = \frac{1}{2\alpha} \int \sqrt{\bar g} \, \bar g^{\mu\nu} \mathcal F_\mu \mathcal F_\nu \, ,
\end{equation}
and the corresponding Faddeev-Popov ghost action,
\begin{equation}
    S_c = \int \sqrt{\bar g} \, \bar c^\mu \frac{\delta \mathcal F_\mu}{\delta h_{\alpha\beta}} D_\alpha c_\beta \, .
\end{equation}
For the results presented in the main text, we employ a harmonic gauge condition, $\alpha=\beta=1$. More results in the Landau limit $\alpha=0$ are presented in the appendix.

\begin{figure}
\centering
\includegraphics[width=\textwidth]{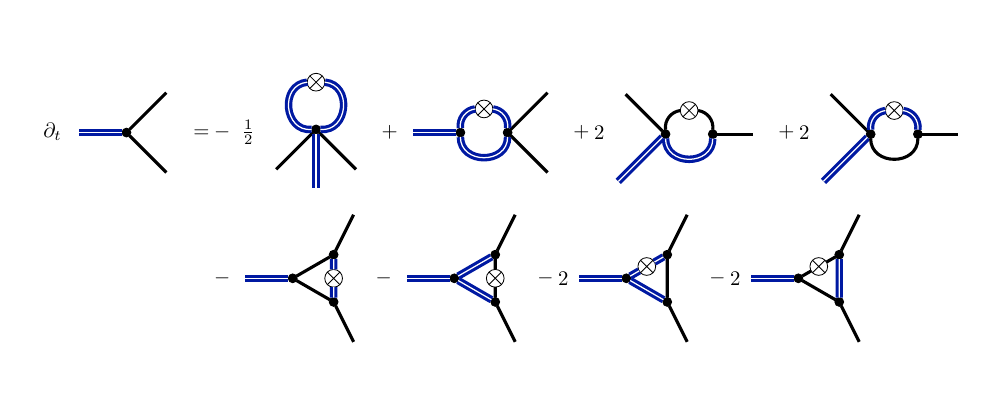}
\caption{\label{fig:diagrep}Diagrammatic representation of the flow equation for the scalar-scalar-graviton vertex. The blue double lines represent gravitons, the black plain lines scalars, and the crossed dot represents the regulator insertion. We refer to the diagrams as ``tadpole'' (first diagram), ``triangle'' (diagrams in second line), and ``swordfish'' (remaining diagrams).}
\end{figure}

For the scalar-scalar-graviton vertex, we encounter a technical complication due to the fact that the functional \ac{RG} deforms the diffeomorphism symmetry via the gauge-fixing and the regularisation procedure. This entails that we have to parameterise a \emph{general} $h\phi\phi$-vertex, without imposing the relations that would be implied by a diffeomorphism-invariant action like \eqref{eq:EA_FF}. More concretely, we have to compute the fluctuation-correlation functions, see~\cite{Pawlowski:2020qer} for a review. To implement the correct \ac{RG} scaling of the vertices, we parameterise the fluctuations of the metric as
\begin{equation}\label{eq:metricpert}
    g_{\mu\nu} = \delta_{\mu\nu} + \sqrt{32\pi G_{N,k} Z_h(-\partial^2)} \, h_{\mu\nu} \, .
\end{equation}
In this, $Z_h$ is a uniform momentum-dependent wave function of the graviton, and we have used a flat Euclidean background metric $\bar g=\delta$. The corresponding anomalous dimension, $\eta_h=-\partial_t \ln Z_h$, has been computed in~\cite{Christiansen:2014raa, Knorr:2021niv}. We also dress the scalar with its wave function $Z_\phi$. This suggests the ansatz for the $h\phi\phi$-part of the effective action
\begin{equation}\label{eq:ansatzFF}
\begin{aligned}
    &\Gamma_k|_{(h\phi\phi)} = \sqrt{32\pi G_{N,k}} \int \text{d}^4x \, \sqrt{Z_h(-\partial_1^2) Z_\phi(-\partial_2^2) Z_\phi(-\partial_3^2)} \, \Bigg[ \frac{1}{2} \Fgk(-\partial_1^2,-\partial_2^2,-\partial_3^2) h \phi_1 \phi_2 \\
    &- \frac{1}{2} \Fdsqk(-\partial_1^2,-\partial_2^2,-\partial_3^2) h^{\mu\nu} (\partial_\mu \partial_\nu \phi_1) \phi_2 - \frac{1}{2} \Fddk(-\partial_1^2,-\partial_2^2,-\partial_3^2) h^{\mu\nu} (\partial_\mu \phi_1) ( \partial_\nu \phi_2) \Bigg] \, .
\end{aligned}
\end{equation}
One can show that this is a complete basis for this vertex, see \Cref{app:basis}. Note that $\Fgk$ and $\Fddk$ are symmetric in their second and third argument, whereas $\Fdsqk$ is not. Moreover, in a shift-symmetric theory, there are further constraints on $\Fgk$ and $\Fdsqk$ for specific momentum configurations. In this work, we compute all three form factors with their complete momentum dependence for the first time. For this, we evaluate the diagrams shown in \Cref{fig:diagrep} in a flat background, so that we can use momentum space techniques. We assign the ingoing momenta $p_h, p_{\phi_1}, p_{\phi_2}$ to the external fields.

Earlier work resolved a reduced momentum dependence of the form factor $\Fg$~\cite{Chiesa:2026tlz}, or the constant part of the vertices~\cite{Eichhorn:2017sok, Laporte:2021kyp, Knorr:2022ilz, Ohta:2025xxo, deBrito:2025nog}. For work on the non-minimal interaction between gravity and scalars without shift symmetry, see also \eg{}~\cite{Narain:2009fy, Henz:2013oxa, Percacci:2015wwa, Labus:2015ska, Henz:2016aoh, Pawlowski:2018ixd, Wetterich:2019zdo, Eichhorn:2020sbo, Sen:2021ffc, Maitiniyazi:2025pou}.

To facilitate the computation, we implement several simplifications. First, we neglect the cosmological constant, and we model the flow of Newton's constant by
\begin{equation}
    G_{N,k} = \frac{G_N}{1 + \frac{G_N k^2}{g_\ast}} \, .
\end{equation}
Here, $g_\ast$ is the fixed point value of the dimensionless Newton's constant, and $G_N=G_{N,k=0}$ is the dimensionful Newton's constant at $k=0$. This approximation models the true scale dependence remarkably well~\cite{Bonanno:2000ep, Fehre:2021eob}.

Second, we neglect the impact of the anomalous dimensions, \ie{}, we set
\begin{equation}
    \eta_h(p^2) = \eta_\phi(p^2) = 0 \, .
\end{equation}
In Landau-deWitt gauge, the scalar anomalous dimension is not generated to leading order~\cite{Meibohm:2015twa}, whereas in the harmonic gauge used in this work, it is small~\cite{Knorr:2019atm}, so we consider this to be a mild approximation.\footnote{Eventually, even in Landau-deWitt gauge, $\eta_\phi$ will be generated by the impact of non-trivial vertex functions.} The graviton anomalous dimension has been computed in~\cite{Christiansen:2014raa, Knorr:2021niv}. We neglect it here, together with the form factors $F_{R,C}$, for technical simplicity. Since the full graviton propagator seems to qualitatively behave like a dressed classical graviton~\cite{Bonanno:2021squ, Fehre:2021eob, Pawlowski:2025etp, Knorr:2026jcg, Assant:2026dca}, we expect sub-leading corrections from including it.

Third, we neglect the feedback of the non-trivial form factors into their own flow equations. Rather, we model all propagators and vertices by their form stemming from the Einstein-Hilbert action with a free scalar field~\eqref{eq:gamma_minimal}, subject to the expansion~\eqref{eq:metricpert}. As a result, the flow equations for the form factors are linear differential equation, instead of non-linear integro-differential equations. This is likely the most severe approximation, and it has to be lifted in future work. From our results, we indeed find that in the high-momentum regime, this approximation breaks down, see the discussion in \Cref{sec:limitations}, and the feedback has to be taken into account self-consistently, as has been done in~\cite{Chiesa:2026tlz}.

To analyse the flow equations, we introduce dimensionless quantities. The dimensionless Newton's constant is $g_k=G_{N,k}k^{2}$. For the external momenta, we use
\begin{equation}
    \left[\tilde p_i^2\right] = \left[p_i^2\right]/k^2 \, ,
\end{equation}
whereas for the form factors,
\begin{equation}
    F_{X,k}\argsofF = k^{d_X} f_{X,k}\argsoff \, ,
\end{equation}
with $d_g=2$ and $d_{\partial\partial\phi}=d_{\partial\phi\partial\phi}=0$.
Using the projectors given in \Cref{app:basis}, we can then extract their beta functions as
\begin{equation}\label{eq:FFflowequations}
\begin{aligned}
    &\dot f_{X,k}\argsoff + \bigg[ d_X + \frac{1}{2} \frac{\dot g_k-2g_k}{g_k} \bigg] \, f_{X,k}\argsoff \\
    &\hspace{2.5cm} - 2 \mathbf{\tilde p}^2\cdot\nabla_{\mathbf{\tilde p}^2} \, f_{X,k}\argsoff = \text{flow}_X\argsoff \, .
\end{aligned}
\end{equation}
Here, we defined the right-hand sides via
\begin{equation}
    \text{flow}_X = \frac{1}{\sqrt{32\pi g_k}\sqrt{Z_h\left(\phsqdimless\right) Z_\phi\left(\ponesqdimless\right) Z_\phi\left(\ptwosqdimless\right)}} \Pi_X^{\mu\nu} \frac{\delta^3 \dot \Gamma}{\delta h^{\mu\nu}(p_h)\delta\phi(p_{\phi_1})\delta\phi(p_{\phi_2})} \, .
\end{equation}
The explicit expressions for all $\text{flow}_X$ follow from the evaluation of the diagrams in \Cref{fig:diagrep}.\footnote{In passing, we note that in our approximation, the tadpole diagram only contributes to the flow of $\fg$ and $\fdd$, but not to that of $\fdsq$.}
Note that in our approximation, all flow terms are linear in $g_k$. It will prove useful to thus write
\begin{equation}
    \text{flow}_X\argsoff = g_k \, \widetilde{\text{flow}}_X\argsoff \, ,
\end{equation}
where $\widetilde{\text{flow}}$ is independent of $g_k$.

Last but not least, for the regulator we follow the standard strategy
\begin{equation}
    \mathfrak R_k(p^2) = \Gamma^{(0,2)}_k(p^2) \frac{\mathcal R_k(p^2/k^2)}{p^2} \, ,
\end{equation}
with the regulator shape being the same in all sectors (graviton and scalar), namely
\begin{equation}
    \mathcal R_k(z) = k^2 \frac{z}{e^z-1} \, .
\end{equation}
This choice, together with the approximations above, allows us to do most of the loop integrals analytically. In particular, all diagrams in the first line in \Cref{fig:diagrep} can be evaluated analytically. For the tadpole, this is straightforward. For the three swordfish diagrams, the reason is that the effective non-trivial momentum dependence is one-dimensional -- only a single external momentum appears in the propagators. Correspondingly, the non-trivial parts of the loop integral are the radial direction and one angular direction, which can both be performed analytically by using Feynman parameters. By contrast, the triangle diagrams in the second line do not reduce to an effective one-dimensional momentum dependence, and thus they do not admit a straightforward analytical integration. Nevertheless, we are able to cast the expressions for both the fixed-point form and the \ac{IR} form factors into a single Feynman-parameter integral which can be evaluated numerically in a straightforward manner. The remaining integrals can also be expanded asymptotically through a matched-asymptotic, or expansion-by-regions analysis~\cite{Smirnov:2021dkb}. We report the expressions for the fixed-point and physical form factors in a supplemental Mathematica notebook. For the computation itself, we employed the \textit{xAct} package~\cite{Martin-Garcia:2007bqa,Brizuela:2008ra,Martin-Garcia:2008ysv,Nutma:2013zea}.

\section{Results}
\label{sec:results}

In this section, we discuss our results. We start with the formal solution to the differential equations \eqref{eq:FFflowequations}. We then discuss the form factors at the fixed point, followed by the physical results at $k=0$. The latter includes a discussion for both small and large momenta as well as the limitations of our approximation. While the formal solution to the flow equation, \Cref{sec:formal_sol}, applies to an arbitrary gauge choice, all explicit analytical and numerical results presented here have been obtained in harmonic gauge. Some numerical results in Landau gauge are provided in \Cref{app:landau}.

\subsection{Formal solution to the flow equations}\label{sec:formal_sol}

We now derive the formal solution to the flow equations \eqref{eq:FFflowequations}. We only discuss the equation for $\fgk$ in detail -- the solution for the other two equations follows in a similar manner.

The general solution to \eqref{eq:FFflowequations} for $X=g$ reads
\begin{equation}\label{eq:gensolfg}
    \fgk\argsoff = -\frac{G_N k^2 g_\ast}{2} \sqrt{g_\ast + G_N k^2} \int_0^1 \text{d}\alpha \frac{\widetilde{\text{flow}}_g\left(\alpha \phsqdimless, \alpha \ponesqdimless,\alpha \ptwosqdimless\right)}{\left( \alpha (\alpha g_\ast + G_N k^2) \right)^{3/2}} \, .
\end{equation}
Here, we neglected the possibility to add an arbitrary homogenous solution, since we do not want to generate any non-trivial form factors if the flow were to vanish. The remaining freedom under this condition would be to have arbitrary integration limits. These are fixed by requiring a well-behaved fixed point solution and a finite limit $k\to0$.\footnote{Note that these ambiguities are likely a truncation artefact, as they are strictly related to the linearity of our equations. We expect no ambiguities once the non-linearities of the feedback of the form factors are taken into account.} The integral is finite, since $\widetilde{\text{flow}}_g$ vanishes linearly at vanishing argument due to the shift symmetry of the ansatz used to derive vertices and propagators.

From the solution \eqref{eq:gensolfg}, we can first derive the fixed point solution upon taking $k\to\infty$. We find
\begin{equation}\label{eq:fg_FP}
    f_{g\,\ast}\argsoff = -\frac{g_\ast}{2} \int_0^1 \text{d}\alpha \frac{\widetilde{\text{flow}}_g\left(\alpha \phsqdimless, \alpha \ponesqdimless,\alpha \ptwosqdimless\right)}{\alpha^{3/2}} \, .
\end{equation}
To derive the physical form factors in the limit $k\to0$, we first switch back to dimensionful quantities. Explicitly, we have
\begin{equation}\label{eq:Fgfull}
    \Fgk\argsofF = -\frac{G_N k^4 g_\ast}{2} \sqrt{g_\ast + G_N k^2} \int_0^1 \text{d}\alpha \frac{\widetilde{\text{flow}}_g\left(\alpha \frac{\phsq}{k^2},\alpha \frac{\ponesq}{k^2},\alpha \frac{\ptwosq}{k^2}\right)}{\left( \alpha (\alpha g_\ast + G_N k^2) \right)^{3/2}} \, .
\end{equation}
Now we rescale the integration variable, $\alpha \mapsto \frac{G_N k^2}{g_\ast}\alpha$, which allows us to take the limit $k\to0$. The result is
\begin{equation}
    \Fg\argsofF = -\frac{g_\ast^2}{2G_N} \int_0^\infty \text{d}\alpha \, \frac{\widetilde{\text{flow}}_g\left(\alpha \frac{G_N \phsq}{g_\ast}, \alpha \frac{G_N \ponesq}{g_\ast}, \alpha \frac{G_N \ptwosq}{g_\ast}\right)}{\alpha^{3/2} (1+\alpha)^{3/2}} \, .
\end{equation}
The exact same derivation applies to the other form factors, with minimal changes due to their differing mass dimension. Their fixed-point solution reads
\begin{equation}\label{eq:fddd2_FP}
    f_{\partial\partial\phi/\partial\phi\partial\phi \, \ast}\argsoff = -\frac{g_\ast}{2} \int_0^1 \text{d}\alpha \frac{\widetilde{\text{flow}}_{\partial\partial\phi/\partial\phi\partial\phi}\left(\alpha \phsqdimless, \alpha \ponesqdimless,\alpha \ptwosqdimless\right)}{\alpha^{1/2}} \, ,
\end{equation}
and the $k\to0$ limit gives
\begin{equation}
    F_{\partial\partial\phi/\partial\phi\partial\phi}\argsofF = -\frac{g_\ast}{2} \int_0^\infty\!\!\! \text{d}\alpha \frac{\widetilde{\text{flow}}_{\partial\partial\phi/\partial\phi\partial\phi}\left(\alpha \frac{G_N \phsq}{g_\ast}, \alpha \frac{G_N \ponesq}{g_\ast}, \alpha \frac{G_N \ptwosq}{g_\ast}\right)}{\alpha^{1/2} (1+\alpha)^{3/2}} \, .
\end{equation}
These expressions form the basis for our explicit results below. Note that for both the fixed-point functions and the \ac{IR} form factors, the $\alpha$-integral can be commuted with any remaining Feynman-parameter integral that is still contained within $\widetilde{\text{flow}}_X$, and subsequently performed analytically.

Before we discuss these, let us repeat two observations first made in~\cite{Knorr:2026vax}. From \eqref{eq:Fgfull}, we see that a derivative expansion does not commute with the limit $k\to0$, and thus the former gives rise to spurious divergences in $k$ in this limit~\cite{Basile:2021krr, Knorr:2022ilz, Baldazzi:2023pep, Knorr:2024yiu, Eichhorn:2024wba, DelPorro:2025fiu}. We also see how the scale $k$ is transmuted into the Planck scale along the flow -- the physical form factors only depend on the physical momenta multiplied by Newton's constant, as they must.

\subsection{Fixed-point form factors}

We first analyse the fixed-point form factors, given by the integrals~\eqref{eq:fg_FP} and \eqref{eq:fddd2_FP}. For plotting purposes, it is convenient to use two squared momenta and the cosine of their angle instead of the three squared momenta. This is because the former have a fixed domain, whereas the latter fulfil a triangle inequality. We choose $\ponesqdimless, \ptwosqdimless$ and
\begin{equation}
    x_{12} = \frac{\phsqdimless-\ponesqdimless-\ptwosqdimless}{2\sqrt{\ponesqdimless \ptwosqdimless}} \, .
\end{equation}
For the fixed point, we focus purely on the Euclidean regime, meaning that the two squared momenta are non-negative, whereas $x_{12}\in[-1,1]$.

\begin{figure}
        \includegraphics[width=0.485\textwidth]{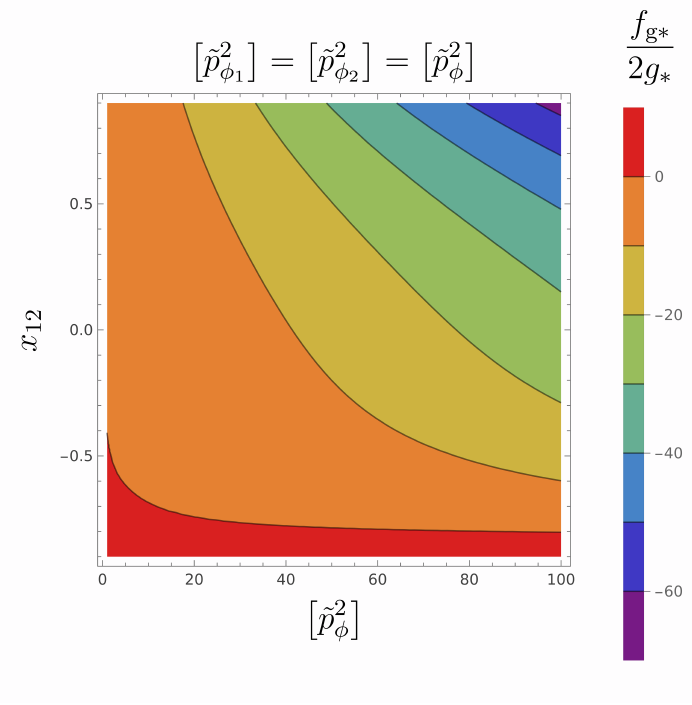} \hfill \includegraphics[width=0.485\textwidth]{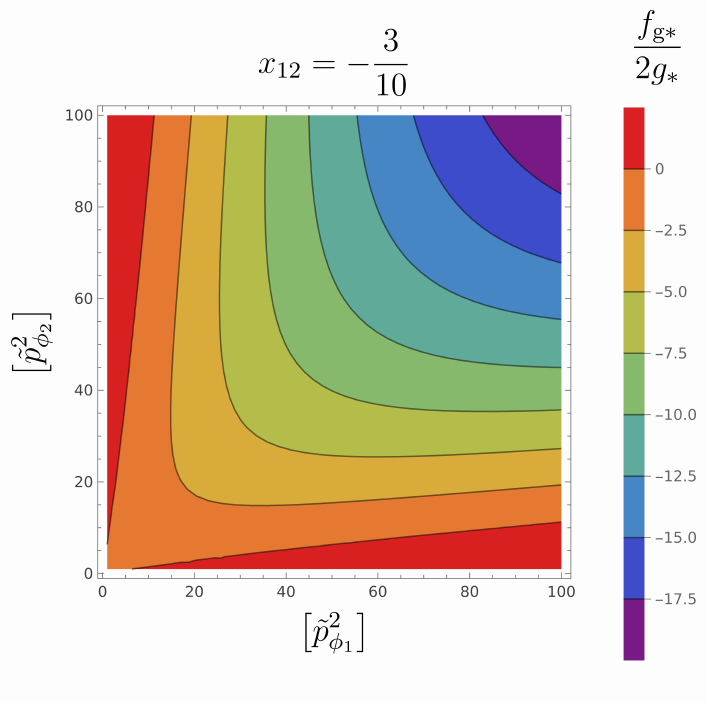}
        
        \includegraphics[width=0.485\textwidth]{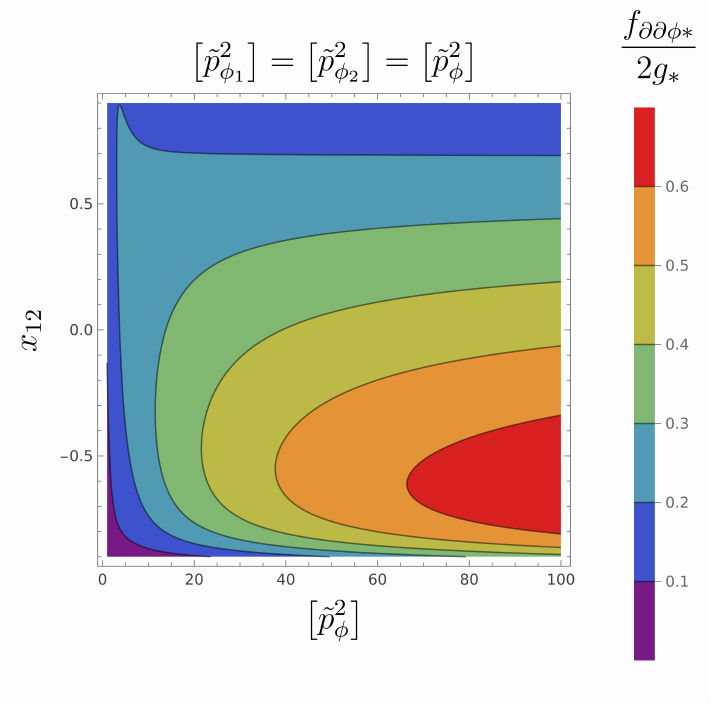} \hfill \includegraphics[width=0.485\textwidth]{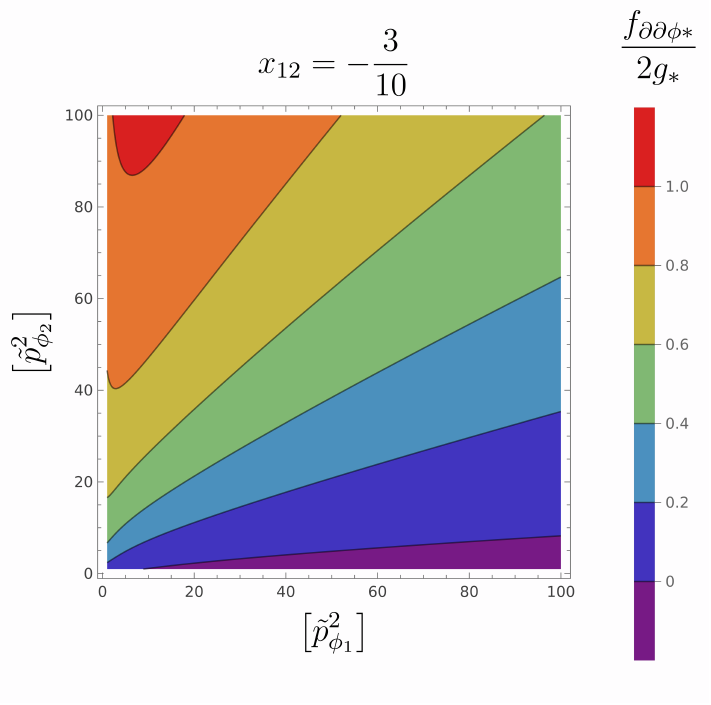}
        
        \includegraphics[width=0.485\textwidth]{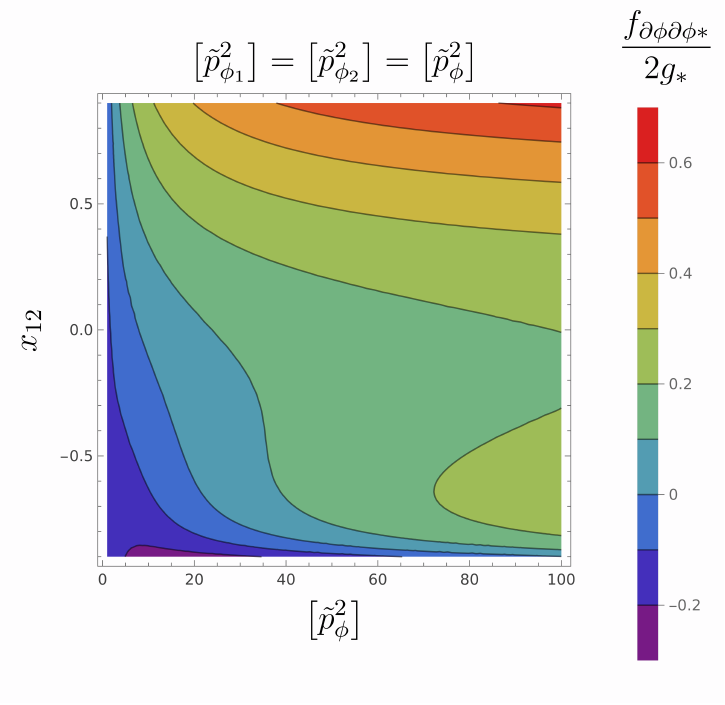} \hfill \includegraphics[width=0.485\textwidth]{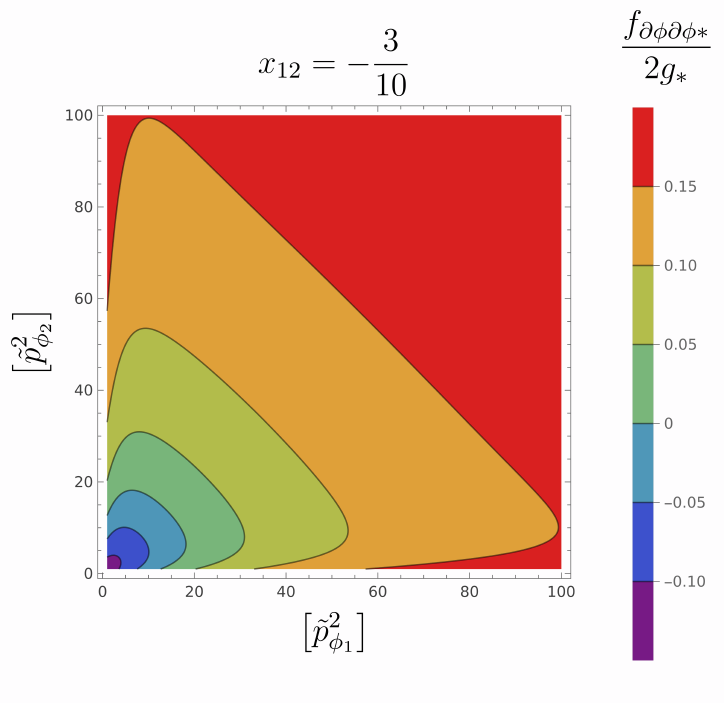}
        \caption{Contour plots of the Euclidean fixed-point form factors, for equal squared scalar momenta (left column), and fixed angle $x_{12}=-3/10$ between the scalar momenta (right column). The latter clearly shows the symmetry under exchange of the two scalar momenta for \fgFP{} and \fddFP{}, whereas no such symmetry applies to \fdsqFP{}.}
        \label{cplotFP}
\end{figure}

We display all form factors on two different cuts through the parameter space in \Cref{cplotFP}, once for equal squared scalar momenta, and once for a fixed angle between the scalar momenta. Overall, we see that the absolute value of \fgFP{} is larger than that of the other two form factors. This is however due to the differing mass dimensions of the form factors -- accounting for this and dividing \fgFP{} by $p^2$, we find that over a large range of momenta, all form factors are of roughly equal size. The angular dependence of \fgFP{} shows a clear tilt in one direction, whereas the angular dependence of the others is more complicated.

For small momenta, we can expand the form factors and find
\begin{subequations}
\begin{align}
    f_{g\ast}\argsoff &\sim -\frac{g_\ast}{3\pi} \left( \phsqdimless - \ponesqdimless - \ptwosqdimless \right) \, , \\
    f_{\partial\partial\phi\ast}\argsoff &\sim \frac{g_\ast}{3\pi} \left( \phsqdimless - \ponesqdimless + 2 \ptwosqdimless \right) \, , \\
    f_{\partial\phi\partial\phi\ast}\argsoff &\sim -\frac{g_\ast}{\pi} + \frac{g_\ast}{18\pi} \left( 2\phsqdimless + \ponesqdimless + \ptwosqdimless \right) \, , \\
    &\hspace{6cm} \text{as }\left[\tilde p_i^2\right]\to0 \, . \notag
\end{align}
\end{subequations}
We note that \fgFP{} and \fdsqFP{} cannot have a constant term due to the shift symmetry of our seed action. These leading-order terms originate from different diagrams for the different form factors. For \fgFP{} we find contributions from all diagrams except the triangle diagrams without a three-graviton vertex. For \fdsqFP{} we find contributions from all diagrams except the tadpole. Lastly, the constant term of \fddFP{} receives contributions from the tadpole and swordfish diagrams, whereas the $p^2$-terms originate from swordfish and triangle diagrams. Recall that due to the tensor structure in \eqref{eq:ansatzFF}, \fg{} and \fdd{} are symmetric in the scalar squared momenta, whereas \fdsq{} is not. A similar expansion for large momenta is difficult due to the remaining Feynman parameter integral in our expressions, but can in principle be carried out through matched-asymptotic analysis.

\subsection{IR form factors}\label{sec:ir_form_factors}

We next discuss the limit $k\to0$, from which we can extract the physical form factors entering the effective action. In our simple setup, for plotting purposes it is convenient to introduce the rescaled dimensionless form factors
\begin{equation}
    \bar F_g = G_N F_g/\left(2 g_\ast\right)^2 \, , \qquad \bar F_{\partial\partial\phi,\partial\phi\partial\phi} = F_{\partial\partial\phi,\partial\phi\partial\phi}/\left(2 g_\ast\right) \, ,
\end{equation}
as well as rescaled dimensionless momenta
\begin{equation}
    \bar p^2 = G_N p^2/\left( 2 g_\ast\right) \, .
\end{equation}
This removes all explicit occurrences of $G_N$ and $g_\ast$.

\begin{figure}
        \includegraphics[width=0.485\textwidth]{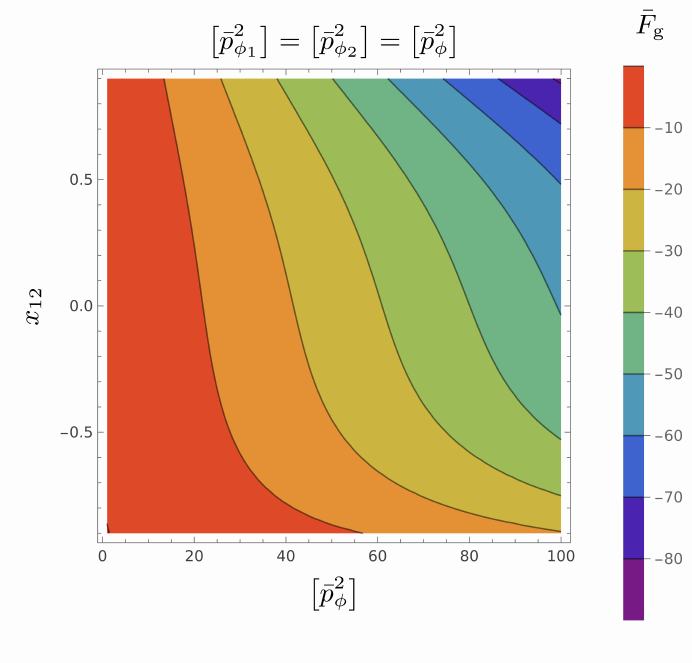} \hfill \includegraphics[width=0.485\textwidth]{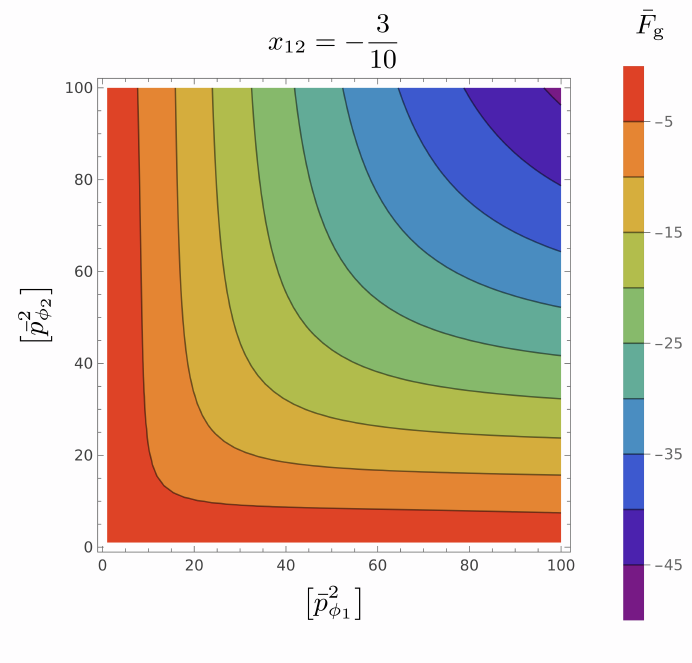}
        
        \includegraphics[width=0.485\textwidth]{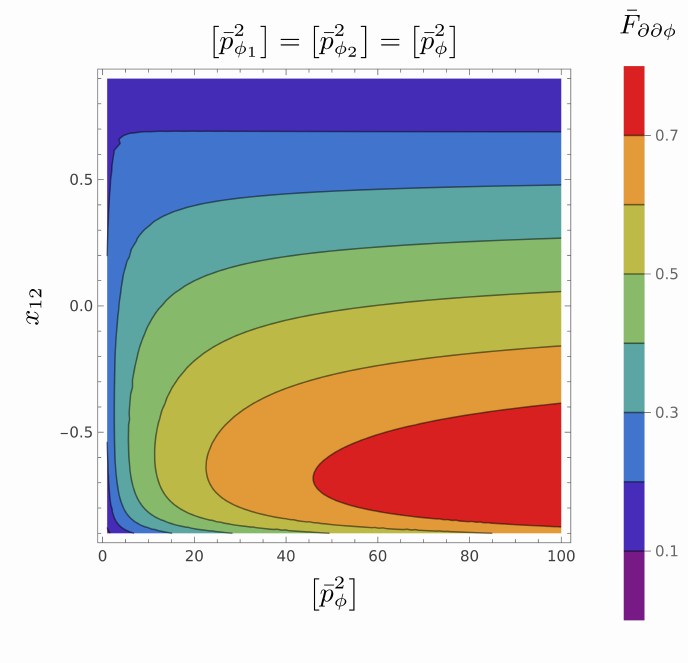} \hfill \includegraphics[width=0.485\textwidth]{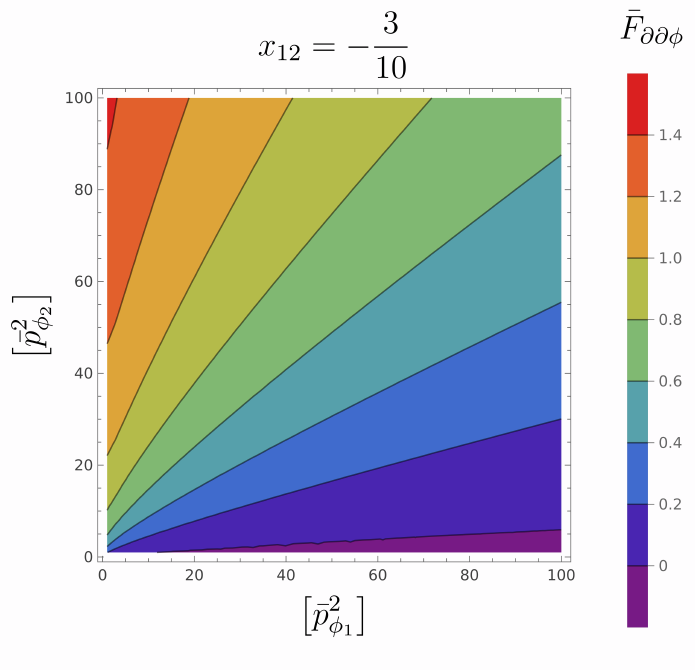}
        
        \includegraphics[width=0.485\textwidth]{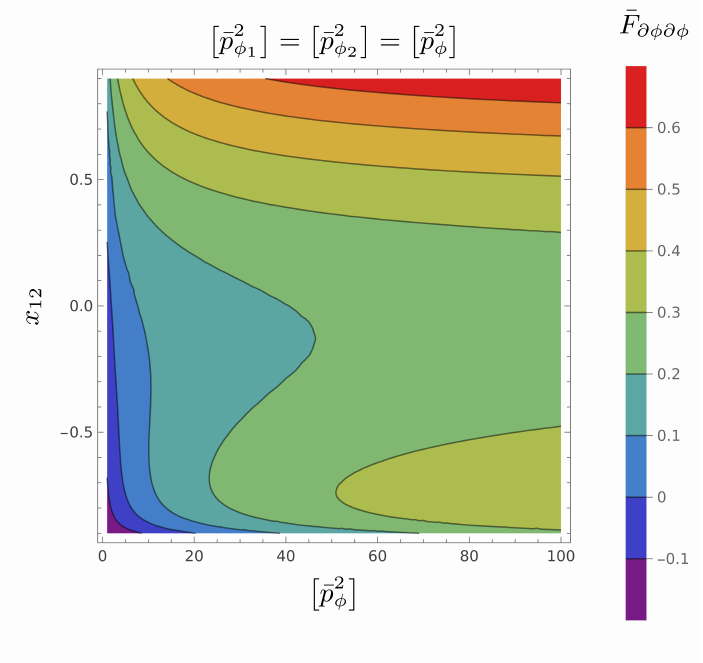} \hfill \includegraphics[width=0.485\textwidth]{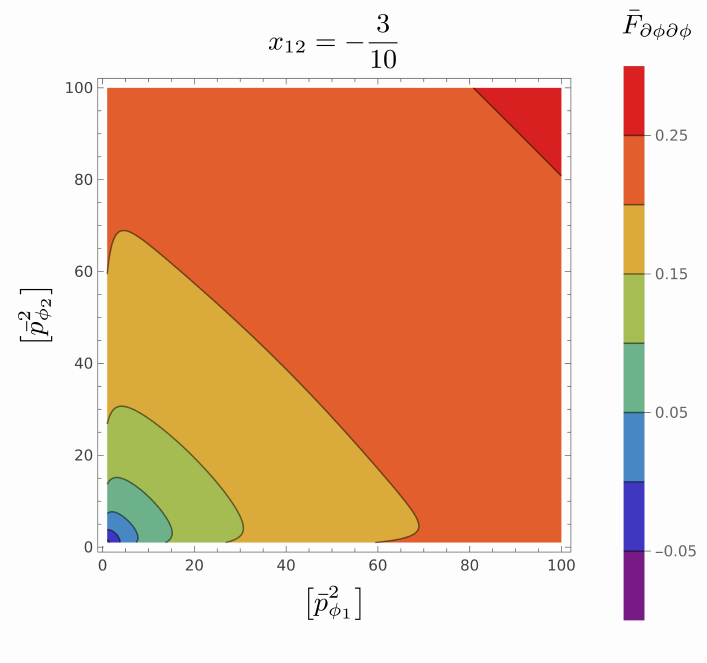}
        \caption{Contour plots of the rescaled dimensionless \ac{IR} form factors, for two different Euclidean configurations: equal squared scalar momenta (left column), and fixed angle $x_{12}=-3/10$ between the scalar momenta (right column). Momenta are rescaled by $g_\ast$ and measured in Planck units, $\bar p^2 = G_N p^2/(2 g_\ast$).}
        \label{cplotIR}
\end{figure}

First, we analyse the Euclidean branch, \ie{}, all squared momenta are non-negative. We show different cuts through the parameter space in \Cref{cplotIR}. Comparing with the fixed-point form factors displayed in \Cref{cplotFP} for the same configurations, we find qualitative similarities for finite squared momenta, but there are some key differences in specific limits. For example, while the fixed-point form factors remain finite at the origin, the \ac{IR} form factors develop logarithmic terms that are connected to the physical branch cuts of the theory. Like at the fixed point, we find that the absolute value of the form factor \Fg{} is larger than that of the other two form factors. Again, this is explained by the differing mass dimensions -- the contribution of all form factors to the scattering amplitude accommodates extra powers of momenta, so that all three form factors generally contribute roughly equally. Since there is no overlap of the form factor $\Fdsq$ with the classical kinetic term, we conclude that the non-minimal interactions \emph{cannot} be neglected in Asymptotically Safe Quantum Gravity.

\begin{figure}
        \includegraphics[width=0.485\textwidth]{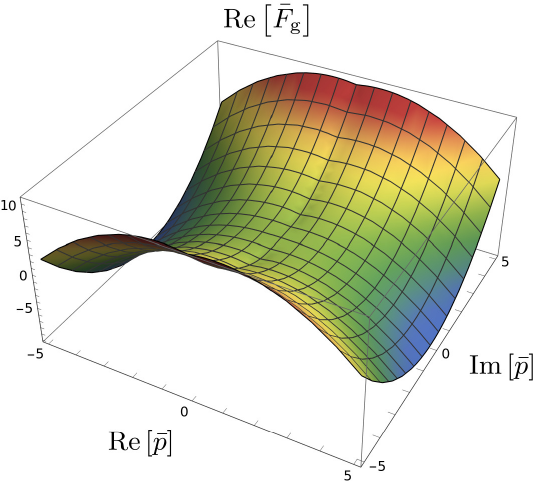} \hfill \includegraphics[width=0.485\textwidth]{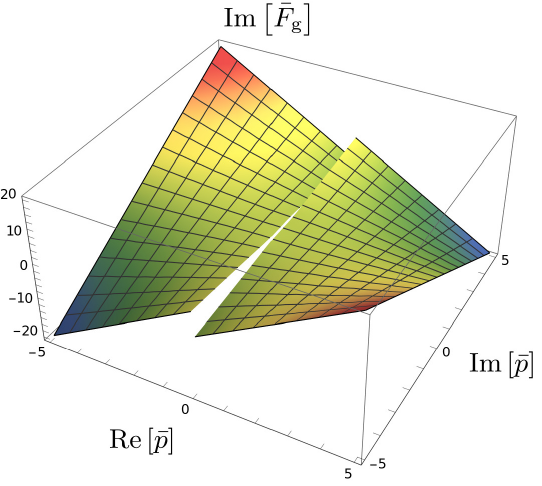} \vspace{0.3cm}\\
        \includegraphics[width=0.485\textwidth]{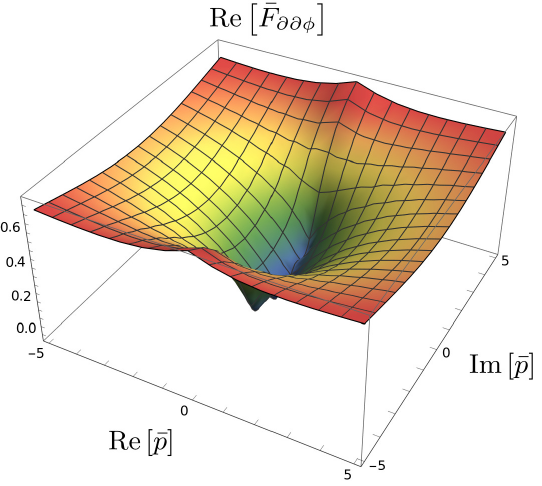} \hfill \includegraphics[width=0.485\textwidth]{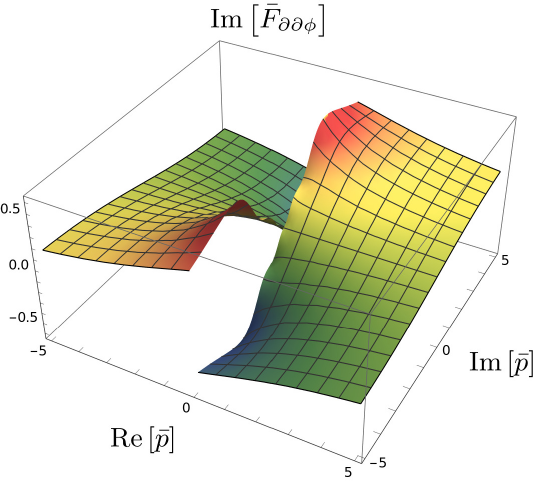} \vspace{0.3cm}\\
        \includegraphics[width=0.485\textwidth]{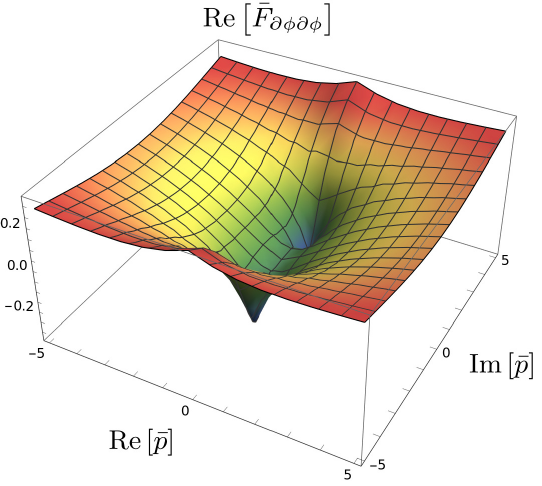} \hfill \includegraphics[width=0.485\textwidth]{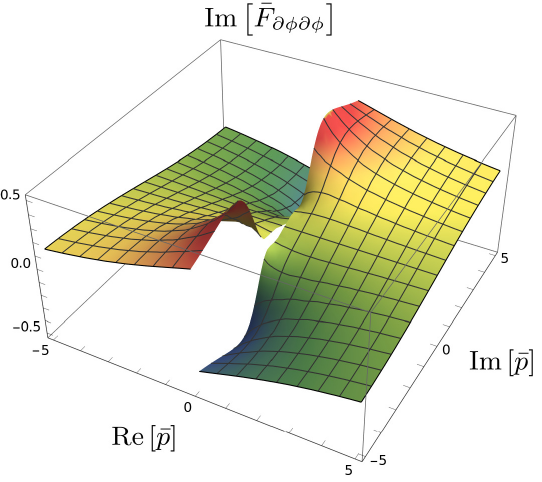}
        \caption{Plots of the real (left panel) and imaginary (right panel) parts of the rescaled \ac{IR} form factors, for the symmetric configuration $\left(\ponesq = \ptwosq = \phsq \equiv \left[p^2\right]\right)$. All panels are plotted as a function of the real and imaginary parts of a rescaled, dimensionless momentum $\left[\bar p\right]=\sqrt{G_N}[p]/\sqrt{2 g_\ast}$.}
        \label{3dplotssym}
\end{figure}

\begin{figure}
        \includegraphics[width=0.485\textwidth]{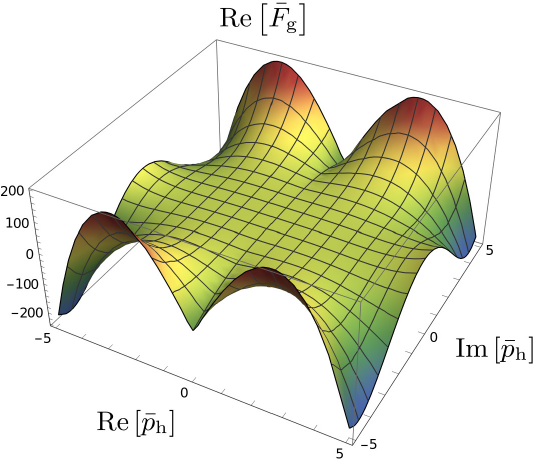} \hfill \includegraphics[width=0.485\textwidth]{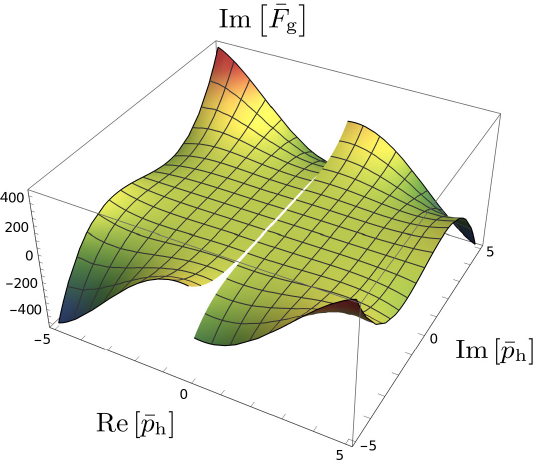} \vspace{0.3cm}\\
        \includegraphics[width=0.485\textwidth]{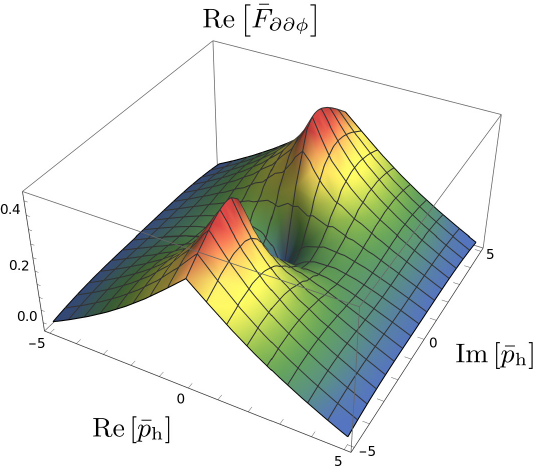} \hfill \includegraphics[width=0.485\textwidth]{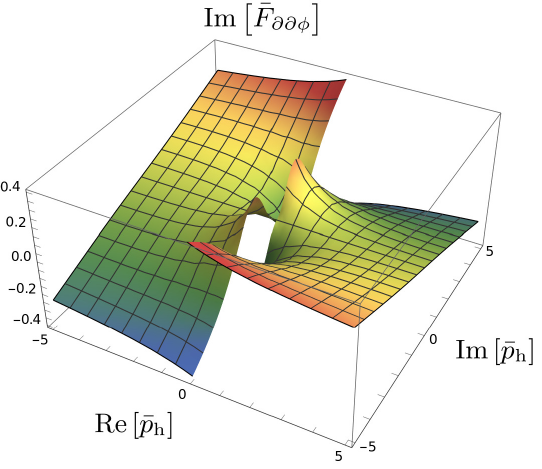} \vspace{0.3cm}\\
        \includegraphics[width=0.485\textwidth]{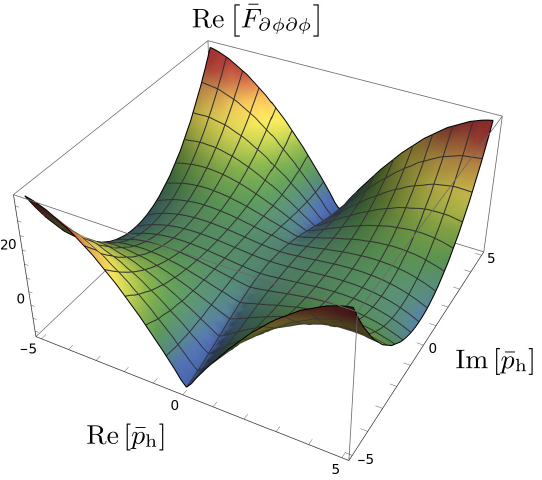} \hfill \includegraphics[width=0.485\textwidth]{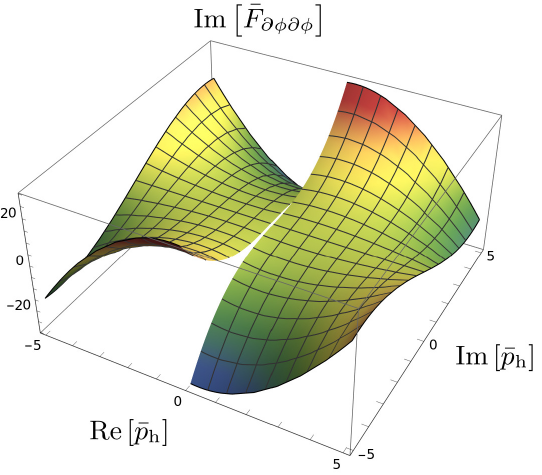}
        \caption{Plots of the real (left panel) and imaginary (right panel) parts of the rescaled \ac{IR} form factors, in the limit relevant for the scalar scattering amplitude, $\ponesq=\ptwosq=0$. We subtracted the soft divergences for \Fg{} and \Fdd{}. All panels are plotted as a function of the real and imaginary parts of the rescaled, dimensionless momentum $\left[\bar p_h \right]=\sqrt{G_N}[p_h]/\sqrt{2 g_\ast}$.}
        \label{3dplotsonshell}
\end{figure}

Second, we discuss the form factors in the complex plane, once again for specific momentum configurations. In \Cref{3dplotssym} we show the symmetric configuration where all squared momenta are equal, whereas in \Cref{3dplotsonshell} we show the on-shell configuration $\ponesq=\ptwosq=0$. For the latter, we subtract the soft divergence, whose precise form is discussed in the next subsection. All real parts are smooth whereas the imaginary parts display a branch cut along the imaginary axis.

Lastly, we give the expansion of the form factors in the on-shell configuration for small and large $\phsq$. In these expressions, the soft divergences as discussed in the next section are again subtracted. For small \phsq{}, we find
\begin{subequations}
\begin{align}
    \Fg\left(\phsq,0,0\right) &\sim -\frac{g_\ast \phsq}{3\pi} + \frac{G_N \phsq^2}{6\pi} \left[ \frac{43}{72} + \gamma - \frac{\pi^2}{4} + \ln \frac{\sqrt{3} G_N \phsq}{16g_\ast} \right] \, , \\
    \Fdsq\left(\phsq,0,0\right) &\sim -\frac{G_N \phsq}{2\pi} \left[ \frac{61}{54} + \gamma + \ln \frac{G_N \phsq}{8g_\ast} \right] \, , \\
    \Fdd\left(\phsq,0,0\right) &\sim -\frac{g_\ast}{\pi} + \frac{G_N \phsq}{6\pi} \left[ \frac{49}{18} - \gamma + \pi^2 - \ln \frac{81 G_N \phsq}{2048 g_\ast} \right] \, , \\
    &\hspace{8cm} \text{as }\phsq\to0 \, , \notag
\end{align}
\end{subequations}
where $\phsq>0$ is understood to correspond to the Euclidean branch, and $\gamma$ is Euler's gamma constant. Note that the coefficients of the leading-order terms of \Fg{} and \Fdd{} agree with those of \fgFP{} and \fddFP{}, but not that of \Fdsq{}. A potential reason is that a putative leading-order term for the latter would also be a constant, which is however forbidden due to shift symmetry. From these expansions, we can extract Wilson coefficients upon fixing a logarithm subtraction scheme, see~\cite{Basile:2021krr, Knorr:2024yiu, DelPorro:2025fiu, Knorr:2026vax, Knorr:2026jcg}. Higher orders can be derived systematically from the full expressions provided in the supplemental Mathematica notebook.

For large \phsq{}, we find
\begin{subequations}\label{eq:FFlargephsq}
\begin{align}
    \Fg\left(\phsq,0,0\right) &\sim \frac{G_N \phsq^2}{16\pi} \Bigg[ \left( \ln \frac{G_N \phsq}{4g_\ast} \right)^2 + 2(\gamma-1)\ln \frac{G_N \phsq}{g_\ast} + 2(5-2\gamma) \ln 2 \notag \\
    &\qquad + \frac{\pi^2}{6} + \gamma(\gamma-2) - 6 \Bigg] - \frac{g_\ast \phsq}{4\pi} \left[ 5\ln \frac{G_N \phsq}{g_\ast} - 24\ln2 + 5\gamma + 6 \right] \notag \\
    &\qquad +\frac{\sqrt{g_\ast^3 \phsq}}{24\sqrt{G_N \pi}} \Bigg[ 84\sqrt{2} - 101\sqrt{3} - 12\pi + 24\ln \left[ \left(3 - 2\sqrt{2}\right)\left(2+\sqrt{3}\right) \right] \Bigg] \notag \\
    &\qquad + \frac{2g_\ast^2}{G_N \pi^2} \ln \frac{G_N \phsq}{g_\ast} + \frac{g_\ast^2}{2\pi G_N} \Bigg[ 9 + 8\gamma - 28\ln2 \Bigg] \, , \\
    \Fdsq\left(\phsq,0,0\right) &\sim -\frac{2g_\ast}{\pi} \left[ \ln \frac{G_N \phsq}{32g_\ast} +\gamma - \frac{11}{4} \right] \, , \\
    \Fdd\left(\phsq,0,0\right) &\sim -\frac{G_N \phsq}{4\pi} \Bigg[ \left( \ln \frac{G_N \phsq}{4g_\ast} \right)^2 + 2(\gamma-1)\ln \frac{G_N \phsq}{g_\ast} + 2(5-2\gamma) \ln 2 \notag \\
    &\qquad + \frac{\pi^2}{6} + \gamma(\gamma-2) - 6 \Bigg] + \frac{g_\ast}{\pi} \left[ 7 \ln \frac{G_N \phsq}{g_\ast} - 32\ln 2 + 7\gamma \right] \, , \\
    &\hspace{9cm} \text{as }\phsq\to\infty \, , \notag
\end{align}
\end{subequations}
up to terms that vanish at infinity. Note that the leading terms of \Fg{} and \Fdd{} are related to each other up to an overall factor of $-\phsq/4$, while at the same overall power in \phsq{}, there is no term in \Fdsq{}. This precise combination of form factors realises a partial restoration of diffeomorphism invariance, as the resulting tensor structure is that of the classical kinetic term, see also the discussion on the soft factor in the next subsection. Again, sub-leading terms can be derived systematically from the provided expressions using matched asymptotics.

\subsubsection{Soft factor}

Our results for the \ac{IR} form factors can be used to evaluate the mediated contribution to the gravitational scattering of scalar fields, cf. the left diagram of \autoref{fig:amp}. For massless external states, we have to Wick-rotate the squared momenta, and then take the limit $\ponesq, \ptwosq \to 0$. This limit, even at the level of the contribution to the scattering amplitude, is obstructed due to soft divergences, as dictated by the Weinberg soft theorem~\cite{Weinberg:1965nx}. It is a highly non-trivial check for our setup whether these divergences are reproduced correctly. This check has two levels: first, divergences should be proportional to a diffeomorphism-invariant operator. This is non-trivial in our setup since physical diffeomorphism symmetry is broken along the flow. Second, both the functional form and the numerical prefactors of the divergences should agree with the soft theorem. This is non-trivial due to our approximations of the \ac{RG} flow, and thus serves as a quality check for the truncation.

We find that our setup reproduces the correct form of the soft eikonal double logarithms as predicted by Weinberg's theorem. Below, we discuss some of the details.

Due to our simple approximation, we can analyse the soft divergences for each form factor on a per diagram basis. Starting with $\Fdsq$, we find soft logarithms from different diagrams, which however cancel when summed up. In other words, the form factor $\Fdsq$ in the on-shell limit is free from all soft divergences. By contrast, both $\Fg$ and $\Fdd$ retain soft divergences, even when all diagrams are summed up. We find in particular that the soft parts are
\begin{equation}\label{eq:soft_div}
    \Fg^\text{soft} = -\frac{\phsq}{4} \Fdd^\text{soft} = -\frac{G_N \phsq^2}{8\pi} \left[ \left( \ln \frac{\ponesq}{\phsq} + \frac{3}{2} \right)\left( \ln \frac{\ptwosq}{\phsq} + \frac{3}{2} \right) - \frac{9}{4} \right] \, .
\end{equation}
Previous numerical results on this vertex focussed on the large-momentum properties~\cite{Chiesa:2026tlz}. In \Cref{app:comparison} we discuss how the soft terms could be identified in such a setup.

Let us put this result into perspective. First, the precise relation between $\Fg$ and $\Fdd$ reflects the diffeomorphism symmetry of our soft divergences: the corresponding tensor structure of the vertex is exactly the one coming from a variation of the classical kinetic term, cf. \eqref{eq:ansatzFF}, in agreement with the soft theorem. Second, we find quadratic and linear logarithms. To leading order, the quadratic logarithms are completely constrained by the soft theorem, including the prefactor. Our result matches this, and thus as claimed above, we reproduce the double eikonal limit. In turn, the linear logarithms are not uniquely determined for the form factors, but only for the full scattering amplitude. This entails that for a complete soft check, the quartic scalar interaction needs to be computed at a similar level of approximation as we did for the three-point vertex, for a first step in this direction see~\cite{Knorr:2026vax}. This reinforces how only the complete scattering amplitude is physical, whereas individual contributions from a subset of diagrams is not. We leave such a complete investigation of the scalar scattering amplitude to future work.

\subsubsection{Limitations of the approximation}\label{sec:limitations}

Let us discuss the regime of validity of our approximation. The fact that we can reproduce the soft double-eikonal divergences indicates that our approximation captures the small momentum regime well. On the other hand, recall that we approximated the vertices entering the flow diagrams in \Cref{fig:diagrep} by their classical form stemming solely from the scalar kinetic term and the Einstein-Hilbert action. Counting powers of momenta in our ansatz for the vertex \eqref{eq:ansatzFF} suggests that this can work qualitatively if
\begin{equation}
    \Fg \propto p^2 \, ,  \qquad \Fdd, \Fdsq \propto 1 \, ,
\end{equation}
as any momentum grows large. If we compare this with the actual scaling of the form factors computed, \eqref{eq:FFlargephsq}, we find that all of our form factors grow faster than this. This supports our earlier assessment that likely the most severe approximation is the neglect of the self-feedback of the form factors. Consequently, scattering amplitudes derived within our approximation might not display physical Asymptotic Safety. This also fits the discussion in~\cite{Chiesa:2026tlz}, which reports that the self-feedback is crucial to obtain a bounded scattering amplitude. We conclude from this that going forward, a combined strategy to resolve vertex functions is most promising: an analytical leading-order estimate as used here delivers information on the \ac{IR}, including the analytical branch cut structure and soft terms, which then informs a fully numerical computation that takes into account the self-feedback as well as the correct branch cut information in the numerical Wick rotation.

\section{Conclusion}
\label{sec:conclusions}

In this work, we have taken significant steps on the way to assess the unitarity and causality of Asymptotically Safe Quantum Gravity. Gravitational scattering amplitudes are a prime target in this endeavour. For the concrete example of the gravitational scattering of scalar fields, we provided an approximation of the scalar-scalar-graviton vertex, including the full momentum dependence in all tensorial channels.

We find that starting from the fixed point, we can derive fully analytic, or simple Feynman-parameter integral, representations for the physical vertex functions. While the computation relies on Euclidean signature, the high degree of analytic control allows a straightforward analytic continuation to Lorentzian momentum configurations. For the on-shell configuration relevant to scattering, we find the expected branch cut structure along the imaginary graviton momentum axis.

A key finding is that, once the differing mass dimensions of the form factors are accounted for, all form factors contribute quantitatively approximately equally. In other words, we find that non-minimal operators \emph{cannot} be neglected in Asymptotically Safe Quantum Gravity.

Our approximation is under control best for small momenta. Here, we find the expected soft double-eikonal divergences predicted by the Weinberg soft theorem~\cite{Weinberg:1965nx}. This is remarkable as along the \ac{RG} flow, our setup breaks diffeomorphism symmetry due to regularisation and gauge-fixing. Nevertheless, the soft divergences computed in our setup are diffeomorphism-invariant and reproduce the correct numerical prefactor.

The high degree of analytic control comes at a price, however. For large momenta, we find that neglecting the self-feedback of the vertex introduces a leading-order error, in line with the findings of a previous numerical computation of the same vertex~\cite{Chiesa:2026tlz}. In such a numerical setup, resolving the soft terms is more challenging though.

This leads us to a clear conclusion for how correlation functions should be computed. Analytic approximations as discussed here and in~\cite{Knorr:2026vax, Knorr:2026jcg} provide information about the analytic structure of correlation functions, which are particularly relevant for \ac{IR} physics. These then inform fully numerical setups in a number of ways: they identify the numerically most important channels (in our case, all of them) as well as where branch cuts are expected. The numerical setup then needs to take into account a self-consistent closure to accommodate the high-momentum limit correctly.

There are some technical barriers which have to be overcome in the future to make our analytic framework applicable to a wider range of theories. First, the specific choice of the regulator shape function that allows the explicit evaluation of threshold integrals only works straightforwardly for gapless theories. A first extension should thus generalise the framework to gapped theories. Second, due to their different kinetic structure, fermions need a separate treatment, and a simple shape function as used here might not straightforwardly lead to similar analytic results.

Our results can be systematically extended to compute other vertex functions. The clear next target is the four-scalar vertex, extending~\cite{Knorr:2026vax} to include all diagrams. Another interesting target is to investigate the emission of a soft graviton by considering five-particle processes. We hope to report on such results soon elsewhere.

\section*{Acknowledgements}

We would like to thank Astrid Eichhorn for collaboration at an early stage of this work, and Angelo P. Chiesa, Astrid Eichhorn, Jan M.~Pawlowski, and Manuel Reichert for discussions. \\

\paragraph{Disclosure of GenAI usage} Codex (model ChatGPT-5.6 Sol Ultra) has been used in \Cref{app:comparison} to design and perform the numerical benchmark, as well as plot its results. All results have been verified by the authors.

% TODO: include funding information
\paragraph{Funding information}

TE acknowledges support by the Deutsche Forschungsgemeinschaft (DFG) under Grant No. 569836646 (GI 328/13-1).

% Authors are required to provide funding information, including relevant agencies and grant numbers with linked author's initials. Correctly-provided data will be linked to funders listed in the \href{https://www.crossref.org/services/funder-registry/}{\sf Fundref registry}.

\begin{appendix}
\numberwithin{equation}{section}

\section{Basis for the scalar-scalar-graviton vertex}\label{app:basis}

In this appendix, we discuss how an arbitrary vertex can be mapped into our basis. For this, we follow the ideas of the classification algorithm in~\cite{Knorr:2022lzn}, and we neglect boundary terms. The relevant part for us is the observation that derivatives that are contracted with other derivatives can be absorbed in a form factor, and thus such terms do not constitute independent basis elements. Consequently, we only have to consider terms with zero or two derivatives on top of the form factor, since the maximum number of free indices coming from the fields are the two graviton indices. The only term without extra derivatives is
\begin{equation}
    \Fg(-\partial_1^2,-\partial_2^2,-\partial_3^2) \, h \, \phi \, \phi \, ,
\end{equation}
which thus constitutes one of our basis elements. For the case of two extra derivatives, we can list all possible ways to distribute them onto the fields. There are four ways to do so,
\begin{equation}
\begin{aligned}
    F_1(-\partial_1^2,-\partial_2^2,-\partial_3^2) \, \partial^\mu \partial^\nu h_{\mu\nu} \, \phi \, \phi \, , \\
    F_2(-\partial_1^2,-\partial_2^2,-\partial_3^2) \, h_{\mu\nu} \, \partial^\mu \partial^\nu \phi \, \phi \, , \\
    F_3(-\partial_1^2,-\partial_2^2,-\partial_3^2) \, \partial^\mu h_{\mu\nu} \, \partial^\nu \phi \, \phi \, , \\
    F_4(-\partial_1^2,-\partial_2^2,-\partial_3^2) \, h_{\mu\nu} \, \partial^\mu \phi \, \partial^\nu \phi \, .
\end{aligned}
\end{equation}
We recognise the second and fourth elements as our choice of basis elements. It is clear that these two elements are independent even under partial integration. For the first element, we find
\begin{equation}
\begin{aligned}
    &\int \, F_1(-\partial_1^2,-\partial_2^2,-\partial_3^2) \, \partial^\mu \partial^\nu h_{\mu\nu} \, \phi \, \phi \\
    = - &\int \, \left[ F_1(-\partial_1^2,-\partial_2^2,-\partial_3^2) \, \partial^\nu h_{\mu\nu} \,  \partial^\mu \phi \, \phi + F_1(-\partial_1^2,-\partial_2^2,-\partial_3^2) \, \partial^\nu h_{\mu\nu} \, \phi \, \partial^\mu \phi \right] \\
    = - &\int \, \left[ F_1(-\partial_1^2,-\partial_2^2,-\partial_3^2) + F_1(-\partial_1^2,-\partial_3^2,-\partial_2^2) \right] \, \partial^\nu h_{\mu\nu} \,  \partial^\mu \phi \, \phi \, .
\end{aligned}
\end{equation}
Consequently, we can absorb such a term by a shift of $F_3$. In turn, we can write the third term as
\begin{equation}
\begin{aligned}
    &\int \, F_3(-\partial_1^2,-\partial_2^2,-\partial_3^2) \, \partial^\mu h_{\mu\nu} \, \partial^\nu \phi \, \phi \\
    = - &\int \, F_3(-\partial_1^2,-\partial_2^2,-\partial_3^2) \, \left[ h_{\mu\nu} \, \partial^\mu \partial^\nu \phi \, \phi + h_{\mu\nu} \, \partial^\mu \phi \, \partial^\nu \phi \right] \, .
\end{aligned}
\end{equation}
Hence, this term can be absorbed by a shift of $F_2$ and $F_4$, which we thus identify as our basis elements.

Since we have a basis for the three-point function, we can also find projectors onto the different basis elements. For this, we go into momentum space. With the abbreviation (sometimes called the K\"{a}ll\'{e}n function)
\begin{equation}
    \reflP = \phsq^2 - 2 \phsq \left(\ponesq + \ptwosq\right) + \left(\ponesq - \ptwosq\right)^2 \, ,
\end{equation}
the projectors read
\begin{align}
    \Pi_g^{\mu\nu} &= \frac{1}{2} \delta^{\mu\nu} + \frac{2}{\reflP} \left[ \Big( \ponesq p_{\phi_2}^\mu p_{\phi_2}^\nu + 2\ptwosq p_{\phi_1}^\mu p_{\phi_1}^\nu \Big) - \Big( \phsq - \ponesq - \ptwosq \Big) p_{\phi_1}^{(\mu} p_{\phi_2}^{\nu)} \right] \, , \\
    \Pi_{\partial\partial\phi}^{\mu\nu} &= \frac{4}{\reflP} \Bigg\{ \ptwosq \delta^{\mu\nu} + \frac{2}{\reflP} \Bigg[ \left(\reflP + 6 \ponesq \ptwosq\right) p_{\phi_2}^\mu p_{\phi_2}^\nu + 6 \ptwosq^2 p_{\phi_1}^\mu p_{\phi_1}^\nu \nonumber \\
    &\hspace{6cm}- 6 \ptwosq \left( \phsq - \ponesq - \ptwosq \right) p_{\phi_1}^{(\mu} p_{\phi_2}^{\nu)} \Bigg] \Bigg\} \, , \\
    \tilde \Pi_{\partial\partial\phi}^{\mu\nu} &= \frac{4}{\reflP} \Bigg\{ \ponesq \delta^{\mu\nu} + \frac{2}{\reflP} \Bigg[ \left(\reflP + 6 \ponesq \ptwosq\right) p_{\phi_1}^\mu p_{\phi_1}^\nu + 6 \ponesq^2 p_{\phi_2}^\mu p_{\phi_2}^\nu \nonumber \\
    &\hspace{6cm} - 6 \ponesq \left( \phsq - \ponesq - \ptwosq \right) p_{\phi_1}^{(\mu} p_{\phi_2}^{\nu)} \Bigg] \Bigg\} \, , \\
    \Pi_{\partial\phi\partial\phi}^{\mu\nu} &= -\frac{2}{\reflP} \Bigg\{ \left(\phsq - \ponesq - \ptwosq\right) \delta^{\mu\nu} + \frac{4}{\reflP} \Bigg[  - 2 \left(\reflP + 6 \ponesq \ptwosq\right) p_{\phi_1}^{(\mu} p_{\phi_2}^{\nu)} \nonumber \\
    &\hspace{3cm}+ 3\left(\phsq - \ponesq - \ptwosq\right) \left( \ponesq p_{\phi_2}^\mu p_{\phi_2}^\nu + 2\ptwosq p_{\phi_1}^\mu p_{\phi_1}^\nu \right) \Bigg] \Bigg\} \, .
\end{align}
The two projectors $\Pi_{\partial\partial\phi}$ and $\tilde\Pi_{\partial\partial\phi}$ both project onto $\Fdsq$, but with a different ordering of its arguments. The two ways to project allow for a consistency check of the flow equations, which is explicitly fulfilled by our computation. With these and our ansatz \eqref{eq:ansatzFF}, we have
\begin{align}
    \frac{\Pi_g^{\mu\nu} \Gamma^{(h\phi\phi)}_{\mu\nu}}{\mathcal N} &= \frac{1}{2} \, \left[ \Fg\left(\phsq,\ponesq,\ptwosq\right) + \Fg\left(\phsq,\ptwosq,\ponesq\right) \right] \, , \\
    \frac{\Pi_{\partial\partial\phi}^{\mu\nu} \Gamma^{(h\phi\phi)}_{\mu\nu}}{\mathcal N} &= \Fdsq\left(\phsq,\ponesq,\ptwosq\right) \, , \\
    \frac{\tilde\Pi_{\partial\partial\phi}^{\mu\nu} \Gamma^{(h\phi\phi)}_{\mu\nu}}{\mathcal N} &= \Fdsq\left(\phsq,\ptwosq,\ponesq\right) \, , \\
    \frac{\Pi_{\partial\phi\partial\phi}^{\mu\nu} \Gamma^{(h\phi\phi)}_{\mu\nu}}{\mathcal N} &= \frac{1}{2} \left[ \Fdd\left(\phsq,\ponesq,\ptwosq\right) + \Fdd\left(\phsq,\ptwosq,\ponesq\right) \right] \, ,
\end{align}
where
\begin{equation}
    \mathcal N = \sqrt{32\pi G_{N,k} Z_h\left(\phsq\right) Z_\phi\left(\ponesq\right) Z_\phi\left(\ptwosq\right)} \, .
\end{equation}

\section{Numerical resolution of soft divergences}\label{app:comparison}

For a self-consistent computation of vertices, numerical methods are indispensable. The scalar-scalar-graviton vertex that we investigated in this work was studied numerically in~\cite{Chiesa:2026tlz}. Within such a setup, it is challenging to identify the soft divergences. In this appendix, we discuss the potential numerical signatures of these soft terms.

For this, we consider a modified version of the soft divergence \eqref{eq:soft_div}, and use the methods of~\cite{Chiesa:2026tlz} to analytically continue simulated data points on the Euclidean branch into the complex plane via a rational function. We find that at a fixed number of Euclidean data points, by construction, the rational functions used to continue the data will give a finite result at the point where the original function diverges due to the logarithm. A systematic increase of data points can then reveal the onset of the logarithm.

More concretely, the reconstruction in~\cite{Chiesa:2026tlz} first fixes a spatial momentum $r=|\mathbf{p}|$, samples Euclidean frequencies $\omega>0$, and then continues that frequency to the mass shell, $\omega=-\mathbf{i}r$. We will apply this method to the soft logarithm. For this, we go to the centre-of-mass frame and choose the two incoming scalar momenta as
\begin{equation}
    p_{\phi_1} = (\omega,\mathbf{p}) \, , \qquad p_{\phi_2} = (\omega,-\mathbf{p}) \, ,
\end{equation}
so that the virtual graviton momentum at the same vertex is
\begin{equation}
    p_h = (-2\omega, \mathbf{0}) \, .
\end{equation}
For Euclidean momenta, this entails
\begin{equation}
    z:= p_{\phi_1,E}^2 = p_{\phi_2,E}^2 = \omega^2 + r^2 \, , \qquad x:= p_{h,E}^2 = 4\omega^2 \, .
\end{equation}
As stated before, the on-shell limit is then $\omega\to-\mathbf{i}r$. For the numerical analysis, it is convenient to introduce
\begin{equation}
    \Delta = 1 + \frac{\omega^2}{r^2} \, , \qquad \omega_\Delta = 0^+-\mathbf{i}r\sqrt{1-\Delta} \, .
\end{equation}
The path $\Delta$ from 1 to 0 specifies a clockwise Wick rotation through the fourth quadrant. Along this path,
\begin{equation}
    x(\omega_\Delta) = -4r^2(1-\Delta) -\mathbf{i}0 \, , \qquad z(\omega_\Delta) = r^2 \Delta-\mathbf{i}0 \, .
\end{equation}
Taking $\Delta\to0^+$ realises the on-shell limit at fixed non-zero momentum transfer $x$.

For the numerical investigation, for convenience we consider a simplified version of the soft term \eqref{eq:soft_div}, namely
\begin{equation}\label{eq:trial}
    D(\omega,\Lambda) = \frac{1}{1+\left(\frac{\omega}{\Lambda}\right)^2} - \left[ \frac{1-\left(\frac{r}{\Lambda}\right)^2}{1+\left(\frac{\omega}{\Lambda}\right)^2} \right]^3 \frac{x(\omega)}{2\pi} \ln z(\omega) \left( \ln z(\omega) + 3 - 2 \ln x(\omega) \right) \, .
\end{equation}
In this, $\Lambda>r$ is an arbitrary scale. We choose this trial function as it imposes
\begin{equation}
    D(0,\Lambda) = 1 \, , \qquad D(\omega,\Lambda) \sim_{\omega\to\infty} \frac{1}{\omega^2} \, ,
\end{equation}
so that the trial function is bounded, and the angular bracket in the limit $\omega\to-\mathbf{i}r$ is the identity, so that the soft logarithms are independent of $\Lambda$. Comparing with \eqref{eq:soft_div}, we have set $G_N=g_\ast=1$.

\begin{figure}[t]
        \centering
        \includegraphics[width=0.6\textwidth]{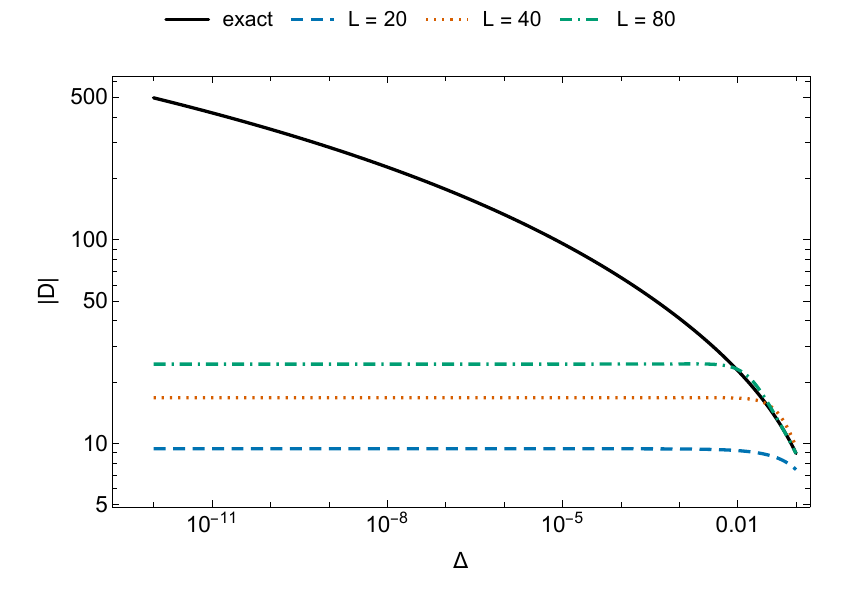}
        \caption{\label{fig:soft_trial}Plot of the absolute value of the trial function \eqref{eq:trial} and its Schlessinger approximants for $L=20,40,80$, as a function of $\Delta$. The limit $\Delta\to0$ probes the on-shell limit, and the trial function develops soft logarithms. In turn, for fixed $L$ the approximants quickly converge to finite values. Systematically increasing $L$ shows that this finite value increases, which we interpret as the numerical sign of the soft logarithm.}
\end{figure}

For the numerical analysis, we choose $\Lambda=6r$ and use units such that $r=1$. We probe 200 Euclidean data points, of which 80 are logarithmically spaced on $[0.02r,r]$, and 120 are linearly spaced on $[1.05r,20r]$. We then select $L\in\{20,40,80\}$ random points and apply the Schlessinger-point method to construct rational approximations. We discard any candidate continuation if it has a pole in the Euclidean region or in the complex frequency quarter-disk $\text{Re} \omega\geq0$, $\text{Im} \omega \leq 0$, $|\omega| \leq r$. Out of the remaining candidates, we retain the one with the smallest relative root-mean-square error on an independent 600-point Euclidean grid.

With this setup, increasing the number of grid points or the sample size $L$ one can reduce the error of the fit in the Euclidean domain as much as desired. For example, for $L=80$ the pointwise relative error in the Euclidean domain is of the order of $10^{-27}$. After analytic continuation, the situation changes however. In particular, we can now study the path for $\Delta$ described above. This is shown in \Cref{fig:soft_trial} for all three choices of $L$. We see that at fixed $L$, all interpolations converge rapidly to a finite value as $\Delta\to0$, which however increases as $L$ increases. The finiteness is intrinsic to the continued-fraction approximant. By contrast, the true trial function \eqref{eq:trial} diverges as the square of a logarithm (plus sub-leading terms),
\begin{equation}
    D(\omega,\Lambda) \sim_{\Delta\to0} \frac{2r^2}{\pi} \left( \ln\Delta \right)^2 \, .
\end{equation}
This shows how one might be able to identify the soft terms in numerical setups like the one in~\cite{Chiesa:2026tlz}: increasing the sample size $L$ shows a non-convergent behaviour after analytic continuation. This can then either be matched to an analytic computation as presented in this work, or be extracted by a fit.

\section{Comparison with Landau-deWitt gauge}\label{app:landau}

\begin{figure}
        \includegraphics[width=0.485\textwidth]{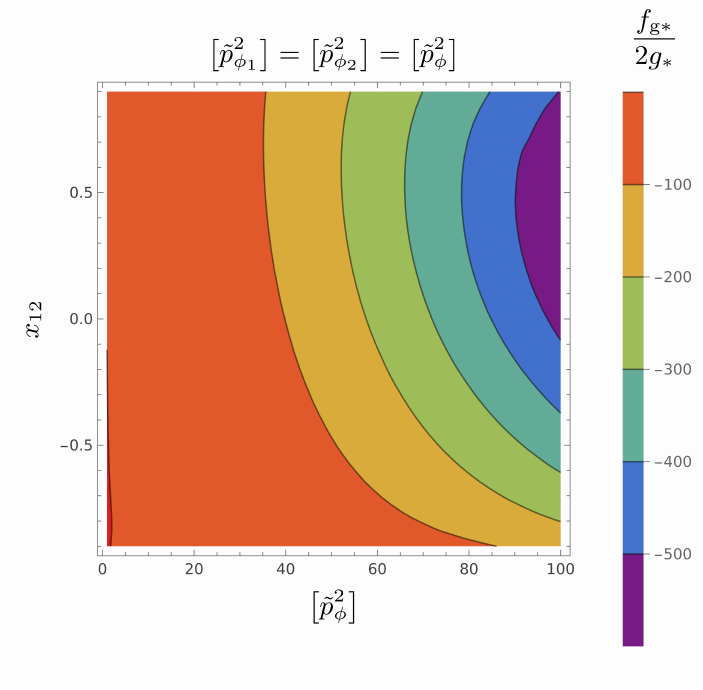} \hfill \includegraphics[width=0.485\textwidth]{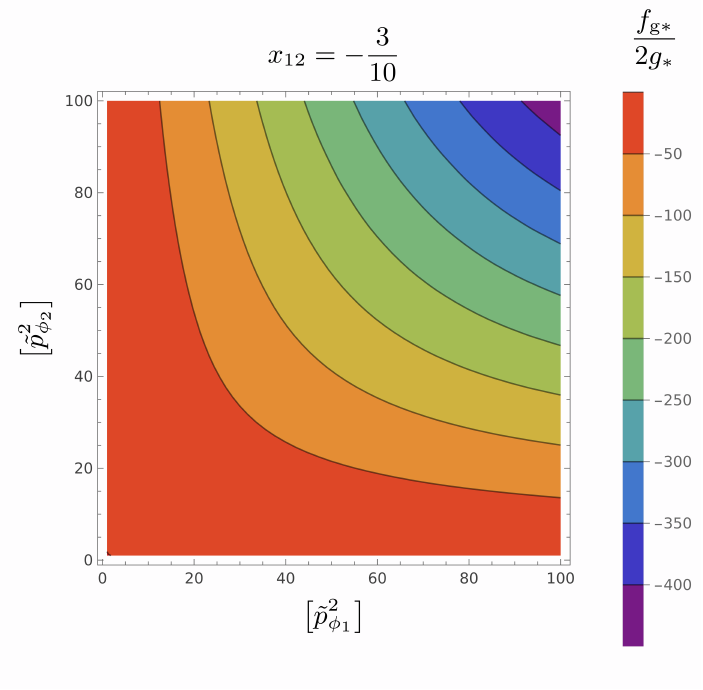}
        
        \includegraphics[width=0.485\textwidth]{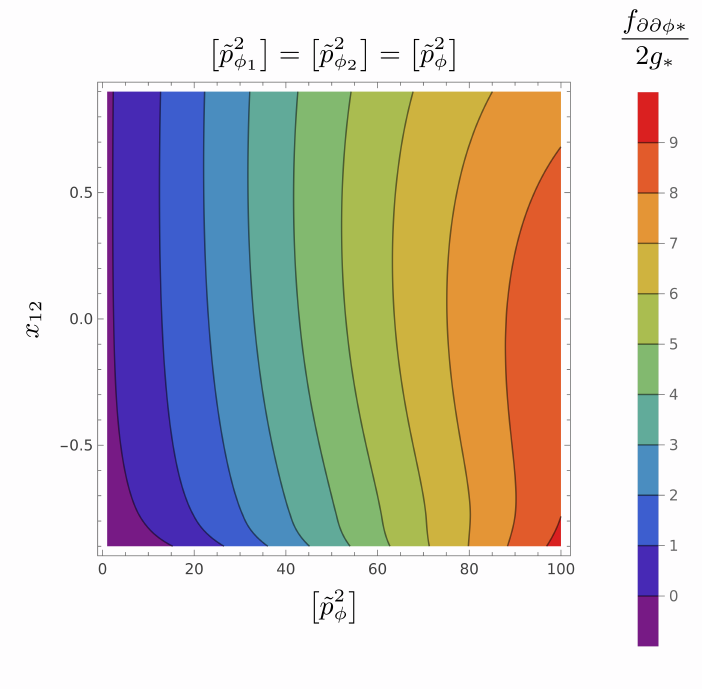} \hfill \includegraphics[width=0.485\textwidth]{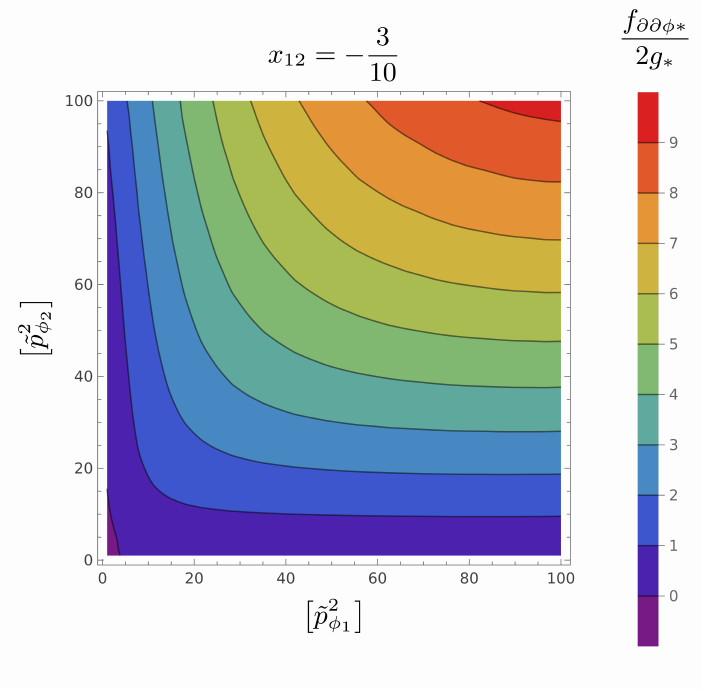}
        
        \includegraphics[width=0.485\textwidth]{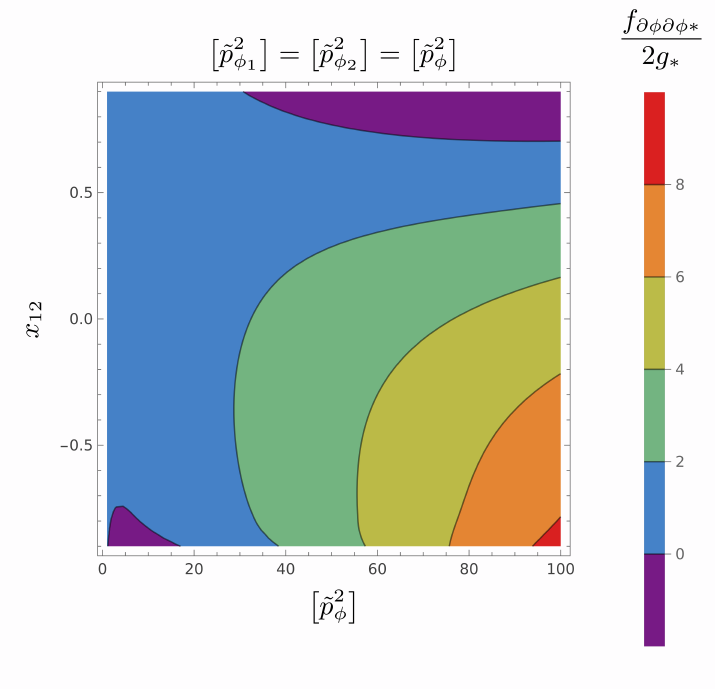} \hfill \includegraphics[width=0.485\textwidth]{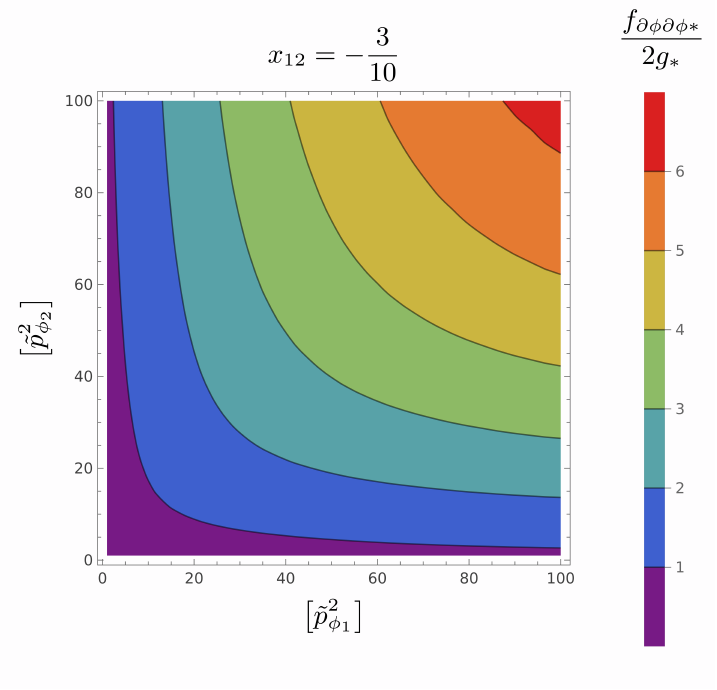}
        \caption{Contour plots of the Euclidean fixed-point form factors in Landau-deWitt gauge, for equal squared scalar momenta (left column), and fixed angle $x_{12}=-3/10$ between the scalar momenta (right column). The latter clearly shows the symmetry under exchange of the two scalar momenta for \fgFP{} and \fddFP{}, whereas no such symmetry applies to \fdsqFP{}.}
        \label{cplotFPL}
\end{figure}

\begin{figure}
        \includegraphics[width=0.485\textwidth]{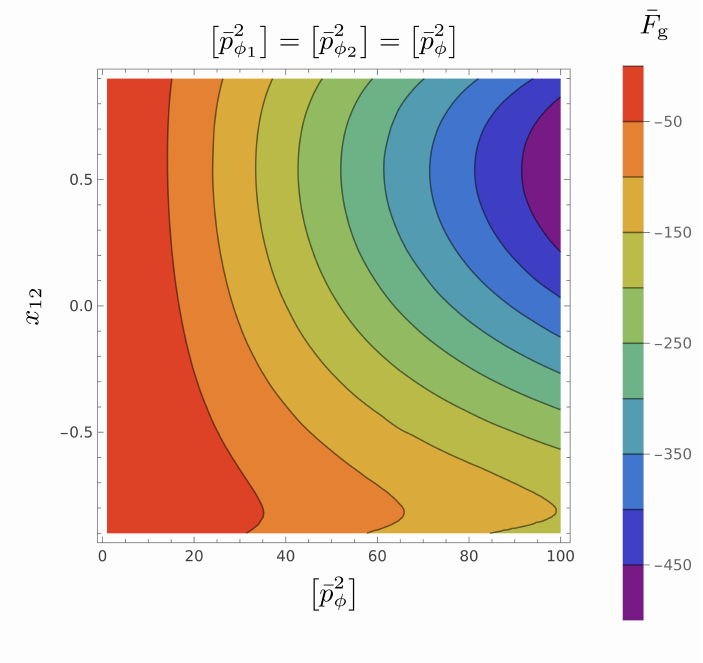} \hfill \includegraphics[width=0.485\textwidth]{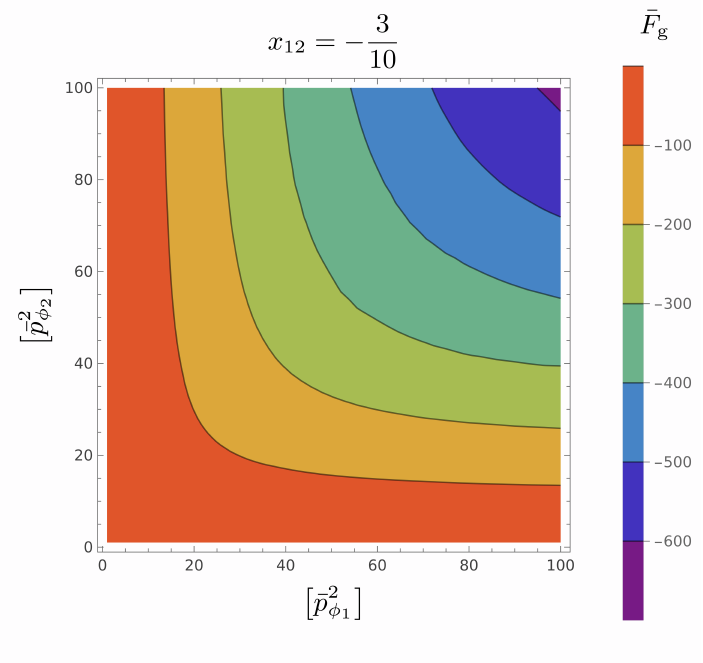}
        
        \includegraphics[width=0.485\textwidth]{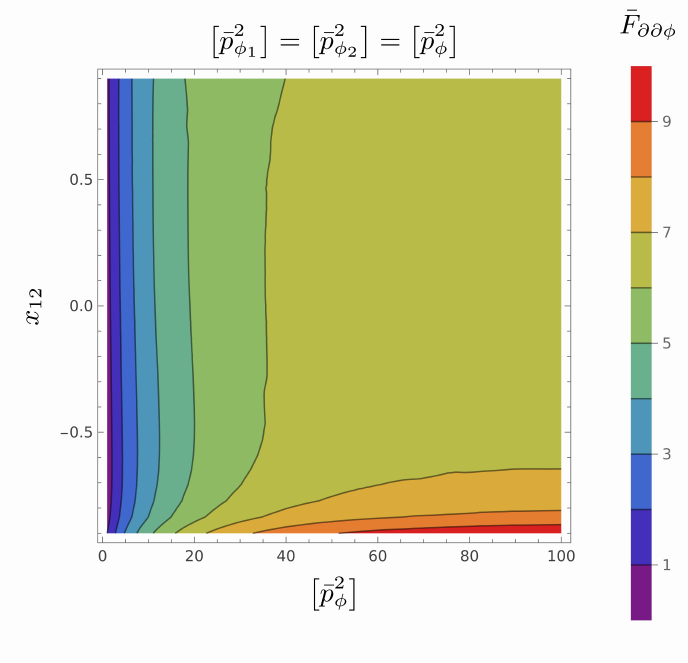} \hfill \includegraphics[width=0.485\textwidth]{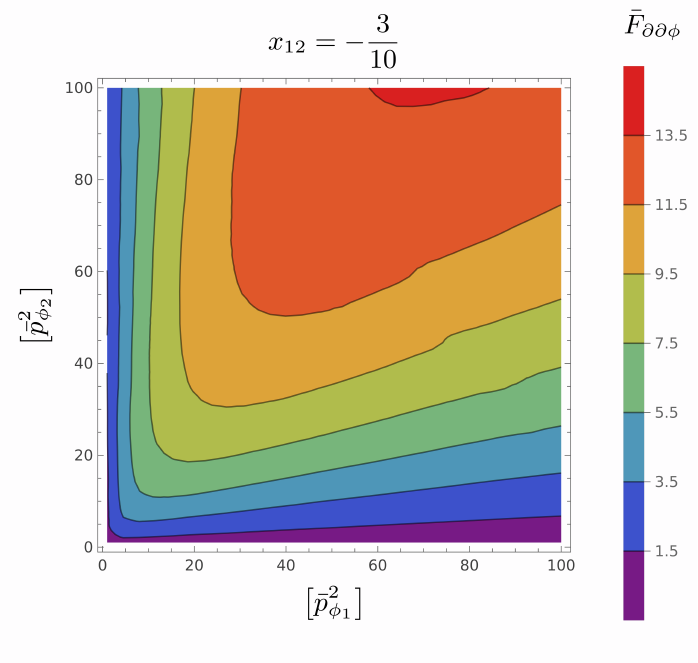}
        
        \includegraphics[width=0.485\textwidth]{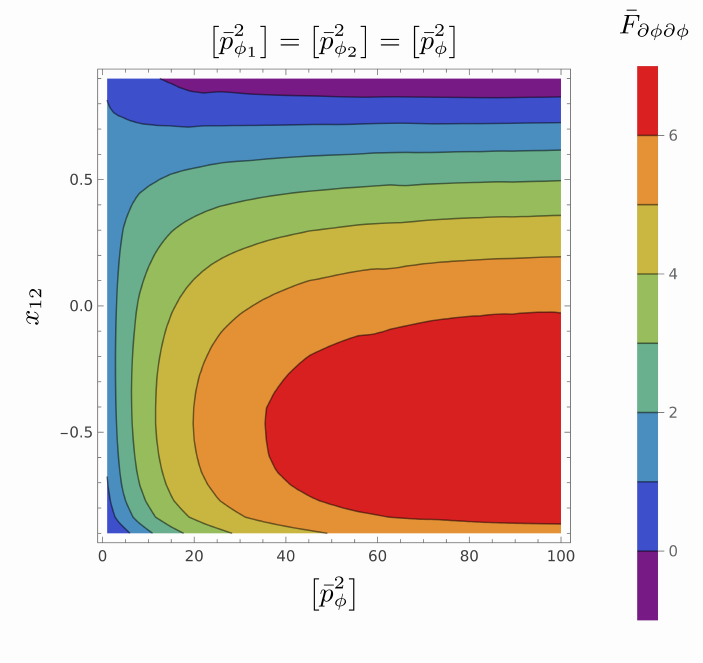} \hfill \includegraphics[width=0.485\textwidth]{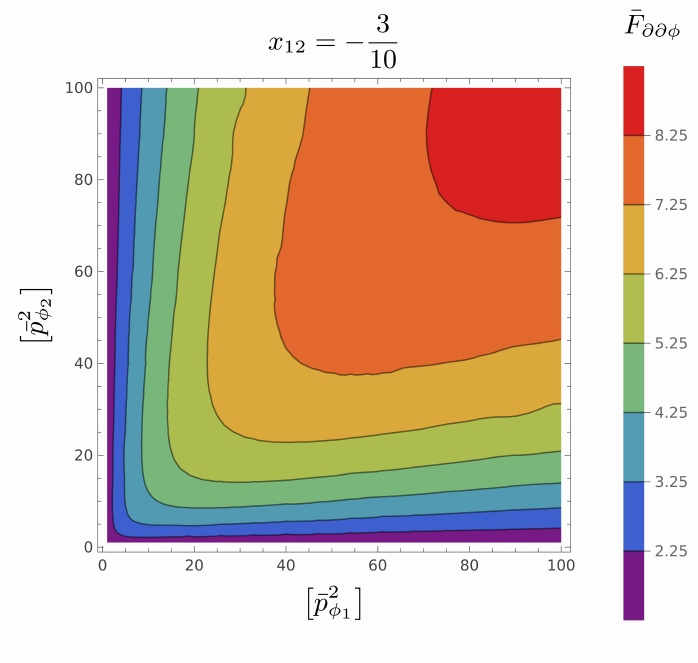}
        \caption{Contour plots of the rescaled dimensionless \ac{IR} form factors in Landau-deWitt gauge, for two different Euclidean configurations: equal squared scalar momenta (left column), and fixed angle $x_{12}=-3/10$ between the scalar momenta (right column). Momenta are rescaled by $g_\ast$ and measured in Planck units, $\bar p^2 = G_N p^2/(2 g_\ast$).}
        \label{cplotIRL}
\end{figure}

To assess the gauge dependence of our results, we compare the \ac{IR} and fixed-point solutions obtained in harmonic gauge with those obtained in Landau-deWitt gauge. While the calculation for the latter choice proceeds in an analogous manner to the one in harmonic gauge, fully analytic expressions are obtained only for the tadpole diagram and the swordfish diagrams. For the triangle diagrams, a subset of the integrations is performed analytically, while the remaining integrals are evaluated numerically. The results were computed on equally-spaced grids and interpolated across the domain. The fixed-point and \ac{IR} form factors in Landau-deWitt gauge are displayed in \Cref{cplotFPL} and \Cref{cplotIRL}, respectively.

Comparing the resulting solutions across both gauge choices, we find some qualitative and even quantitative agreement for \Fg{}, which is similar in both gauge choices at the fixed point and in the \ac{IR}. By contrast, the other form factors exhibit some noticeable differences in their qualitative behaviour between the two gauges. A potential explanation for this is that our truncation only retains the classical scalar kinetic term on the right-hand side of the flow equation, and for a better agreement across different gauge choices, the feedback of the full form factors might be important. The form factors themselves are, on average, approximately one order of magnitude larger in Landau-deWitt gauge than in harmonic gauge. The relative hierarchy among the form factors remains qualitatively similar, with $F_g$ (and $f_{g\ast}$) still being larger than the other form factors. This again is mostly compensated by taking into account the differing mass dimensions. With this, none of the form factors is dominant or subdominant everywhere: each dominates in some region of momentum space, but none does so throughout. Consequently, none of them can be neglected in the computation of scattering amplitudes, as in harmonic gauge (see \Cref{sec:ir_form_factors}). This is a further indication that the self-feedback of the form factors cannot be neglected, cf. \Cref{sec:limitations}.

\end{appendix}

\newpage

\bibliography{bib.bib}

\end{document}